\documentclass[11pt,reqno]{article}
\usepackage{jmlr2e}
\usepackage{fullpage,times,graphicx,amssymb,amsmath,amsfonts,bbm,psfrag,xcolor}

\usepackage{graphicx}
\usepackage{amssymb}
\usepackage{amsmath}
\usepackage{amsfonts}
\usepackage{bbm}
\usepackage{psfrag}
\usepackage{xcolor}
\usepackage{fullpage}
\usepackage{epstopdf}
\usepackage{graphicx}
\usepackage{times}
\usepackage{jmlr2e}
\usepackage{commath}
\usepackage{thmtools}
\usepackage{thm-restate}
\usepackage{times}
\usepackage{graphicx}
\usepackage{natbib}
\usepackage{dirtytalk}
\usepackage{empheq}

\definecolor{ddarkbrown}{rgb}{0.5,0.2,0.05} \definecolor{bbluegray}{rgb}{0.05,0,0.5}

\usepackage[colorlinks,citecolor=bbluegray,linkcolor=ddarkbrown,urlcolor=blue,breaklinks]{hyperref}

\usepackage{subfig,float} 

\usepackage{mathtools}

\usepackage{algorithm,algcompatible,algpseudocode}
\usepackage{multicol,lipsum}
\algnewcommand{\Inputs}[1]{%
	\State \textbf{Inputs: \:}{#1}
}

\algnewcommand{\Output}[1]{%
	\State \textbf{Output: \:}{#1}
}
\algnewcommand{\Initialize}[1]{%
	\State \textbf{Initialize: \:}{#1}
}

\algnewcommand{\IIf}[1]{\State\algorithmicif\ #1\ \algorithmicthen}
\algnewcommand{\EndIIf}{\unskip\ \algorithmicend\ \algorithmicif}

\let \oldsection \section
\renewcommand{\section}{\vspace{3ex plus 1ex}\oldsection}

\newcommand{\BEAS}{\begin{eqnarray*}}
	\newcommand{\EEAS}{\end{eqnarray*}}
\newcommand{\BEA}{\begin{eqnarray}}
\newcommand{\EEA}{\end{eqnarray}}
\newcommand{\mb}{\mathbb}
\newcommand{\BEQ}{\begin{equation}}
\newcommand{\EEQ}{\end{equation}}
\newcommand{\BIT}{\begin{itemize}}
	\newcommand{\EIT}{\end{itemize}}
\newcommand{\BNUM}{\begin{enumerate}}
	\newcommand{\ENUM}{\end{enumerate}}

	\newcommand{\D}{\mathcal{D}}
 \newcommand{\F}{\mathcal{F}}
 
		\newcommand{\U}{\mathcal{U}}
	\newcommand{\Pm}{\mathcal{P}}
	\newcommand{\te}{\theta}
	\newcommand{\mr}{\mathrm}
	\newcommand{\R}{\mathbb{R}}

	\newcommand{\E}{\mathbb{E}}	

	\newcommand{\Sk}{\mathcal{S}}

\newcommand{\M}{\mathcal{M}}

	\newcommand{\ve}{\varepsilon}
\newcommand{\BA}{\begin{array}}
	\newcommand{\EA}{\end{array}}

\newcommand{\mc}{\mathcal}

 \numberwithin{dummy}{section}

\newtheorem{myh}{A.}

\numberwithin{mythm}{section}
\numberwithin{mydef}{section}
\numberwithin{myprop}{section}
\numberwithin{mylem}{section}
\numberwithin{mycor}{section}

\title{Practical Operator Sketching Framework for Accelerating Iterative Data-Driven Solutions in Inverse Problems}

\begin{document}	

	\author{\name Junqi Tang  \email j.tang.2@bham.ac.uk\\
		\addr School of Mathematics,\\ University of Birmingham \\
        \\
       \name Guixian Xu  \email gxx422@student.bham.ac.uk\\
		\addr School of Mathematics,\\ University of Birmingham \\
        \\
        \name Subhadip Mukherjee  \email smukherjee@ece.iitkgp.ac.in\\
		\addr Department of Electronics and Electrical Communication Engineering, \\ Institute of Technology\\ \\
	      \name  Carola-Bibiane Sch\"onlieb  \email cbs31@cam.ac.uk\\
		\addr Department of Applied Mathematics and Theoretical Physics,\\ University of Cambridge 
        }
   
	\editor{}

	\maketitle


\begin{abstract}

We propose a new operator-sketching paradigm for designing efficient iterative data-driven reconstruction (IDR) schemes, e.g. Plug-and-Play algorithms and deep unrolling networks. These IDR schemes are currently the state-of-the-art solutions for imaging inverse problems. However, for high-dimensional imaging tasks, especially X-ray CT and MRI imaging, these IDR schemes typically become inefficient both in terms of computation, due to the need of computing multiple times the high-dimensional forward and adjoint operators. In this work, we explore and propose a universal dimensionality reduction framework for accelerating IDR schemes in solving imaging inverse problems, based on leveraging the sketching techniques from stochastic optimization. Using this framework, we derive a number of accelerated IDR schemes, such as the plug-and-play multi-stage sketched gradient (PnP-MS2G) and sketching-based primal-dual (LSPD and Sk-LSPD) deep unrolling networks. Meanwhile, for fully accelerating PnP schemes when the denoisers are computationally expensive, we provide novel stochastic lazy denoising schemes (Lazy-PnP and Lazy-PnP-EQ), leveraging the ProxSkip scheme in optimization and equivariant image denoisers, which can massively accelerate the PnP algorithms with improved practicality. We provide theoretical analysis for recovery guarantees of instances of the proposed framework.  Our numerical experiments on natural image processing and tomographic image reconstruction demonstrate the remarkable effectiveness of our sketched IDR schemes.\footnote{This paper contains some contents from its short conference version \citep{tang2024iterative}, and some early results/contents from our unpublished technical report \citep{tang2022accelerating}.}

\end{abstract}

\section{Introduction}

Randomized sketching and stochastic first-order optimization methods have become the de-facto techniques in modern data science and machine learning with a wide range of applications \citep{kingma2014adam,johnson2013accelerating,allen2017katyusha, chambolle2018stochastic,chambolle2024stochastic}, due to their remarkable scalability to the size of the optimization problems. The underlying optimization tasks in many applications nowadays are large-scale and high-dimensional by nature, as a consequence of big-data and overparameterized models (for example the deep neural networks).

While well-designed optimization algorithms can enable efficient machine learning, one can, on the other hand, utilize machine learning to develop problem-adapted optimization algorithms via the so-called \say{learning-to-learn} philosophy \citep{andrychowicz2016learning, li2019learning}. Traditionally, the optimization algorithms are designed in a hand-crafted manner, with human-designed choices of rules for computing gradient estimates, step-sizes, etc, for some general class of problems. Noting that although the traditional field of optimization has already obtain lower-bound matching (aka, \say{optimal}) algorithms \citep{woodworth2016tight,lan2012optimal,lan2015optimal} for many important general classes of problems, for specific instances there could be still much room for improvement. For example, a classical way of solving imaging inverse problems would be via minimizing a total-variation regularized least-squares \citep{chambolle2016introduction} with specific measurement operators -- which is a very narrow sub-class of the general class of smooth and convex programs where these \say{optimal} optimization algorithms are developed for. To obtain optimal algorithm adapted for a specific instance of a class, the hand-crafted mathematical design could be totally inadequate, and very often we do not even know a tight lower-bound of it.

One of the highly active areas in modern data science is computational imaging (which is also recognized as low-level computer vision), especially medical imaging including X-ray computed tomography (CT) \citep{buzug2011computed}, magnetic resonance imaging (MRI) \citep{vlaardingerbroek2013magnetic}, and positron emission tomography (PET) \citep{ollinger1997positron}. In such applications, the clinics seek to infer images of patients' inner body from the noisy measurements collected from the imaging devices. Traditionally, dimensionality reduction methods such as stochastic approximation \citep{robbins1951stochastic} and sketching schemes \citep{2015_Pilanci_Randomized,drineas2011faster,avron2013sketching} have been widely applied in solving large-scale imaging problems due to their scalability\citep{kim2015combining,sun2019online,pmlr-v70-tang17a}. Inspired by their successes, in our work, we focus on developing a framework for efficient sketched iterative data-driven algorithms tailored for solving imaging inverse problems. In our framework, we effectively deal with the computational redundancy which are prevalent in all of current state-of-the-art iterative data-driven reconstruction (IDR) schemes including plug-and-play (PnP)/ regularization-by-denoising (RED) schemes and deep unrolling (DU) networks. For example, our framework can accelerate any plug-and-play algorithm via reducing the computational cost of forward/adjoint operators and denoisers.



\subsection{Contributions of this work}

In this work, we make four main contributions:

\begin{itemize}
    \item \textbf{Operator sketching framework for accelerating iterative data-driven reconstruction schemes}
    
    We propose an operator sketching framework for developing computationally efficient iterative data-driven reconstruction (IDR) methods, ranging from plug-and-play algorithms to deep unrolling networks based on sketching scheme which we have tailored for imaging inverse problems for massively improving efficiency. We first derive our operator sketching scheme and obtain a plug-and-play multi-stage sketched gradient (PnP-MS2G) algorithm. Comparing to state-of-the-art approaches such as PnP-SGD \citep{sun2019online}, we can observe numerically significant acceleration. Via applying againt our sketching framework to deep unrolling networks, we develop Learned Stochastic Primal-Dual (LSPD) network, and its accelerated variant Sketched LSPD (SkLSPD) which is further empowered with the sketching approximation of products \citep{2015_Pilanci_Randomized,2016_Pilanci_Iterative,pmlr-v70-tang17a}. Our proposed networks can be viewed as a minibatch and sketched extension of the state-of-the-art unrolling network -- Learned Primal-Dual (LPD) network of \citep{adler2018learned}. Noting that the LPD is a very generic framework which takes most of the existing unrolling schemes as special cases, our acceleration schemes can be extended and applied to many other deep unrolling networks such as the ISTA-Net \citep{zhang2018ista}, ADMM-Net \citep{sun2016deep} and FISTA-Net \citep{xiang2021fista}.

    \item \textbf{Stochastic lazy denoisers for PnP schemes}

    While utilizing operator sketching we mitigate the inefficiency due to high-dimensionality measurement operators, the computational cost of state-of-the-art denoising functions are also not negligible. In our work, we propose first the Lazy-PnP scheme where we further introduce stochastic skipping schemes for mitigating the computational cost of the denoiser, which can be jointly applied with our operator sketching schemes for the ultimate acceleration. Moreover, we leverage recently introduced equivariant PnP priors and propose Lazy-PnP-EQ for improved stability and performance especially when state-of-the-art deep denoising networks are used. Via skipping the calls of the denoiser with high-probability, we can achieve order-of-magnitute acceleration for gradient-based PnP algorithms in scenarios where the computational cost of the denoisers are dominant, such as image superresolution.
    
    
    \item \textbf{Theoretical analysis of our framework}
    
    We provide a theoretical analysis of the basic instance of our framework in accelerating proximal gradient descent and plug-and-play algorithms, from the view-point of stochastic non-convex composite optimization. We provide upper and lower bounds of the estimation errors under standard assumptions, suggesting that our proposed PnP-MS2G have the potential to achieve similar estimation accuracy as its full-batch counterpart.
    
    \item \textbf{Less is more -- the numerical effectiveness of our new plug-and-play methods and deep unrolling methods in imaging inverse problems}
    
    We provide numerical studies on the performance of the proposed new plug-and-play schemes (PnP-MS2G / Lazy-PnP / Lazy-PnP-EQ), showing significantly improved numerical results comparing to the standard plug-and-play schemes in image processing and reconstruction tasks. We also numerically evaluate the performance of our proposed networks on two typical tomographic medical imaging tasks -- low-dose and sparse-view X-ray CT. We compare our LSPD and SkLSPD with the full batch LPD. We found out that our networks achieve competitive image reconstruction accuracy with the LPD, while only requiring a fraction of the computation of it as a free-lunch, in both supervised and unsupervised training settings.
    

\end{itemize}


\section{Background}

\subsection{Imaging Inverse Problems}
In imaging, the measurement systems can be generally expressed as:
\begin{equation}\label{model}
    b = A x^\dagger + w,
\end{equation}
where $x^\dagger \in \mb{R}^d$ denotes the ground truth image (vectorized), and $A \in \mb{R}^{n \times d}$ denotes the forward measurement operator, $w \in \mb{R}^n$ the measurement noise, while $b \in \mb{R}^n$ denotes the measurement data. A classical way to obtain a reasonably good estimate of $x^\dagger$ is to solve a composite optimization problem:
\begin{equation}\label{1st-stage}
    x^\star \in \arg\min_{x \in \mb{R}^d} f_b(Ax) + r(x),
\end{equation}
where data fidelity term $f_b(Ax):= f(b, Ax)$ is typically a convex function (one example would be the least-squares $\|b - Ax\|_2^2$), while $r(x)$ being a regularization term, for example the total-variation (TV) semi-norm, or a learned regularization \citep{mukherjee2021end, mukherjee2020learned}. A classical way of solving the composite optimization problem (\ref{1st-stage}) is via the proximal gradient methods \citep{chambolle2016introduction}, which are based on iterations of gradient descent step on $f$, proximal step on $r$ and momentum step using previous iterates for fast convergence \citep{beck2009fast,nesterov2007gradient}. 

Since modern imaging problems are often in huge-scales, deterministic methods can be very computationally costly since they need to apply the full forward and adjoint operation in each iteration. For scalability, stochastic gradient methods \citep{robbins1951stochastic} and ordered-subset methods \citep{erdogan1999ordered,kim2015combining} are widely applied in real world iterative reconstruction. More recent advanced stochastic variance-reduced gradient methods \citep{xiao2014proximal,defazio2014saga, allen2017katyusha,tang2018rest,driggs2020spring} have also been adopted in some suitable scenarios in imaging for better efficiency \citep{tang2019limitation,tang2020practicality, karimi2016hybrid}.

More recently, deep learning approaches have been adapted in imaging inverse problems, starting from the work of \citep{jin2017deep} on the FBP-ConvNet approach for tomographic reconstruction, and DnCNN \citep{zhang2017beyond} for image denoising. Remarkably, the learned primal-dual (LPD) network \citep{adler2018learned} which mimics the update rule of primal-dual gradient method and utilizes the forward operator and its adjoint within a deep convolutional network, achieves state-of-the-art results and outperforms primal-only unrolling approaches. Despite the excellent performance, the computation of the learned primal-dual method is significantly larger than direct approaches such as FBP-ConvNet. 

\subsection{Iterative Data-Driven Reconstruction}
In this section we introduce the notion of iterative data-driven reconstruction (IDR) algorithms which we will be exploring in this work. The current dominating IDR schemes can be summarized under to catagories: the plug-and-play (PnP) algorithms and deep unrolling networks.

\subsubsection{Plug-and-Play algorithms}

Iterative reconstruction algorithms have become ubiquitous for solving imaging inverse problems such as image deblurring/inpainting/superresolution and tomographic image reconstruction (for example X-ray CT, MRI and PET, etc). Due to their strengths in delivering data-consistent and robust reconstruction, these iterative solvers, especially when combined with advanced image priors \citep{dabov2007image, zhang2017beyond, tachella2020neural} in a \say{plug-and-play} (PnP) manner \citep{egiazarian2007compressed,venkatakrishnan2013plug, romano2017little, reehorst2018regularization}, can still thrive in current era where deep neural networks \citep{jin2017deep} have been successfully adopted in all these problems.

While these classical convex regularization approaches provide theoretically tractable solutions for inverse problems, they have been significantly outperformed by the PnP priors, constructed by advanced image denoisers or deep neural networks. The very first PnP algorithm (probably not very well-known) is actually proposed by \citep{egiazarian2007compressed}, which is a PnP stochastic approximation algorithm with BM3D as the denoiser. The PnP-ADMM of \citep{venkatakrishnan2013plug} and PnP-LBFGS of \citep{tan2024provably} extend the classical methods ADMM and L-BFGS, replacing the proximal operator with the denoiser and have been widely applied in solving inverse problems since then. Then a very similar approach named regularization-by-denoising (RED) has been proposed by \citep{romano2017little,reehorst2018regularization}, which explicitly construct the regularization term using the denoiser and provide improved convenience in parameter tuning. Since a strong link between PnP and RED is established in \citep{cohen2021regularization} under the RED-PRO unifying framework, in this work we refer both plug-and-play and regularization-by-denoising as \say{PnP} for simplicity. Although we mostly focusing on PnP schemes in this work, our framework is obviously also applicable to RED.

For large-scale problems, the PnP-ADMM and PnP-PGD may require long computational time to obtain a good estimate. The PnP-SGD \citep{sun2019online} and stochastic PnP-ADMM methods \citep{tang2020fast,sun2020scalable} can provide significant acceleration compared to the deterministic PnP-ADMM/PnP-LBFGS methods. In this work we propose a generic acceleration of PnP gradient methods using dimensionality-reduction/sketching in the image-space. Moreover, we propose two enhanced acceleration schemes which deal with the computational complexity on the denoiser, via leveraging stochastic skipping of proximal operators \citep{papoutsellis2024we} in optimization, and equivariant denoising schemes for PnP algorithms \citep{terris2024equivariant} for stable application of deep denoising networks in PnP.

\subsubsection{Deep unrolling}
Now we start by presenting the background of the deep unrolling networks, starting from the primal-dual gradient-based optimization algorithm. It is well-known that, if the loss function $f_b(\cdot)$ is convex and lower semi-continuous, we can reformulate the original objective function (\ref{1st-stage}) to a saddle-point problem:
\begin{equation}\label{saddle}
    [x^\star, y^\star] = \min_{x} \max_{y} \{r(x) + \langle Ax, y \rangle - f_b^*(y)\},
\end{equation}
where $f_b^*(\cdot)$ is the Fenchel conjugate of $f_b(\cdot)$:
\begin{equation}
    f_b^*(y) := \sup_{h} \{\langle h, y \rangle - f_b(h)\}
\end{equation}
The saddle-point problem (\ref{saddle}) can be efficiently solved by the primal-dual hybrid gradient (PDHG) method \citep{chambolle2011first}, which is also known as the Chambolle-Pock algorithm in the optimization literature. The PDHG method for solving the saddle-point problem obeys the following updating rule:
 \begin{eqnarray*}
 && \mathrm{\textbf{Primal-Dual Hybrid Gradient (PDHG)}} \\&& - \mathrm{Initialize}\ x_0, \Bar{x}_0 \in \mb{R}^d \ y_0 \in \mb{R}^n\\
 &&\mathrm{For} \ \ \ k = 0, 1, 2,...,  K\\
&&\left\lfloor
\begin{array}{l}
y_{k+1} = \mathrm{prox}_{\sigma f_b^*} (y_k + \sigma A \Bar{x}_k);\\
x_{k+1} = \mathrm{prox}_{\tau r} (x_k - \tau A^T y_{k+1});\\
\Bar{x}_{k + 1} = x_{k+1} + \beta (x_{k+1} - x_{k});
\end{array}
\right.
 \end{eqnarray*}
The PDHG algorithm takes alternatively the gradients regarding the primal variable $x$ and dual variable $y$ and performs the updates. In practice, it is often more desirable to reformulate the primal problem (\ref{1st-stage}) to the primal-dual form (\ref{saddle}), especially when the loss function $f$ is non-smooth (or when the Lipschitz constant of the gradient is large).

Currently most successful deep networks in imaging would be the unrolling schemes \citep{gregor2010learning} inspired by the gradient-based optimization algorithms leveraging the knowledge of the physical models. The state-of-the-art unrolling scheme -- learned primal-dual network of \citep{adler2018learned} is based on unfolding the iteration of PDHG by replacing the proximal operators $\mathrm{prox}_{\sigma f^\star}(\cdot)$ and $\mathrm{prox}_{\tau g}(\cdot)$ with multilayer convolutional neural networks $\Pm_{\te_p}(\cdot)$ and $\D_{\te_d}(\cdot)$, with sets of parameters $\te_p$ and $\te_d$,  applied on the both primal and dual spaces. The step sizes at each steps are also set to be trainable. The learned primal-dual with $K$ iterations can be written as the following, where the learnable paramters are $\{\te_p^k, \te_d^k, \tau_k, \sigma_k\}_{k=0}^{K-1}$:
\begin{eqnarray*}
 && \mathrm{\textbf{Learned Primal-Dual (LPD)}} \\&&- \mathrm{Initialize}\ x_0 \in \mb{R}^d \ y_0 \in \mb{R}^n\\
 &&\mathrm{For} \ \ \ k = 0, 1, 2,...,  K-1\\
&&\left\lfloor
\begin{array}{l}
y_{k+1} = \D_{\te_d^k} (y_k, \sigma_k, A {x}_k , b);\\
x_{k+1} = \Pm_{\te_p^k}(x_k, \tau_k, A^T y_{k+1});\\
\end{array}
\right.
\end{eqnarray*}
When the primal and the dual CNNs are kept fixed across the layers of LPD, it has the potential to learn both the data-fidelity and the regularizer (albeit one might need additional constraints on the CNNs to ensure that they are valid proximal operators). This makes the LPD parametrization more powerful than a learned proximal-gradient network (with only a primal CNN), which can only learn the regularization functional. The capability of learning the data-fidelity term can be particularly useful when the noise distribution is unknown and one does not have a clear analytical choice for the fidelity term. 


We choose the LPD as our backbone because it is a very generic framework which covers all existing gradient-based unrolling schemes and plug-and-play algorithms as special cases. For instance, if we choose the dual subnetworks of LPD to be a simple subtraction $Ax_k - b$, we can recover Learned ISTA/FISTA.

\section{Iterative Operator Sketching Framework}

We propose an operator sketching framework based on dimensionality reduction on both the data-dimensional $n$ and parameter dimension $d$, constructing a much smaller proxy operator to replace the role of full operator in the iterative reconstruction.

Our framework performs sketching in both image domain (of dimension $d$) and data domain (of dimension $n$). For ease of illustration, we use the least-squares objective and linear forward operator here without loss of generality. For a given forward operator $A \in \R^{n \times d}$, we can often find a low dimensional proxy $A_s \in \R^{n \times m_0}$ discretized on a reduced image dimension $m_0 < d$ such that $Ax \approx A_s\Sk(x)$, where $\Sk(\cdot): \R^d \rightarrow \R^{m_0}$ ($m_0 < d$) is a sketching/downsampling operator. Furthermore, we can also perform random sketching $M(\cdot): \R^n \rightarrow \R^{m}$ ($m<n$) on the measurement/data domain, which corresponds to stochastic approximation \citep{robbins1951stochastic}. One typical choice of this sketching operator $M$ is the subsampling sketch -- uniformly sampled minibatch of $I_{n \times n}$ \citep{2015_Pilanci_Randomized}, which is suitable for inverse problems. For the image domain sketching operator $\Sk$, we found that off-the-shelf down-sampling algorithms such as the bi-cubic interpolation suffice in our framework. Now we can summarize this double-sketching as follows:
 \begin{equation}\label{sko}
 \begin{aligned}
      \|b - Ax\|_2^2 &\approx \|b - A_s\Sk(x)\|_2^2  \\&\propto\mathbf{E}_M \|Mb - MA_s\Sk(x) \|_2^2.
 \end{aligned}
 \end{equation}
Instead of using standalone data-domain sketches \citep{2015_Pilanci_Randomized}, our double-sketching framework is more effective in terms of dimensionality reduction and can be applied to generically accelerate PnP methods and also deep unrolling networks. When using the sketched loss in \eqref{sko}, we can have an approximate data fit that can be efficiently optimized by SGD \citep{robbins1951stochastic} or its variance-reduced variants \citep{johnson2013accelerating}. To recover the same reconstruction quality as the original program, we can adjust the image-domain sketch size $m_0$ stage-wise in a coarse-to-fine manner.

We now further introduce our framework first under the context of plug-and-play algorithms:

\subsection{Doubly-Sketched PnP}

In this section we present our multi-stage sketched gradient framework PnP-MS2G. The sketching techniques have been widely applied in large-scale optimization especially the least-squares problems \citep{pilanci2017newton,2016_Pilanci_Iterative,2015_Pilanci_Randomized,tang2017exploiting, pmlr-v70-tang17a}. However, the author has found that such data-domain sketching methods are not efficient in imaging inverse problems, since very often the forward operator is relatively sparse, and even the most efficient sparse Johnson-Lindenstrauss transform \citep{woodruff2014sketching} cannot provide significant computational gain here since the sketched operator typically has similar sparsity as the full operator. If we use subsampling sketch which is the only practical data-domain sketch, the performance is similar or worse than SGD methods in imaging inverse problems.

Instead of using data-domain sketches, we propose to perform sketching in image-domain which appears to be much more effective and can be applied to generically accelerate PnP proximal gradient methods.

\subsubsection{Algorithmic Framework}
Suppose the original objective reads:
\begin{equation}\label{obj_f}
    x^\star \in \arg\min_{x \in \mc{M}} f(b, Ax),
\end{equation}
where $\mc{M}$ can be some implicit non-convex constraint set constructed by the denoiser (we write this for the ease of presentation), then our sketched objective can be generally expressed as:
\begin{equation}\label{obj_s}
    x^\star \in \arg\min_{x \in \mc{M}} f(b, A_s \Sk(x)),
\end{equation}
where $\Sk(\cdot): \R^d \rightarrow \R^m$ ($m < d$) being the sketching/downsampling operator, while $A_s \in \R^{n \times m}$ is the forward operator discretized on the reduced image space. We found that such a scheme provides a remarkably efficient approximation of the solution. We present our PnP-MS2G framework in Algorithm 1, where we denote $\D$ as the denoiser, $\Sk$ as the sketching operator, and $\U$ as the upsampling operator.

\begin{algorithm}[t]\label{AA}
   \caption{--- Plug-and-Play with Multi-Stage Sketched Gradients (PnP-MS2G)}
\begin{algorithmic}

   \State  {\bfseries Initialization:} $x_0 \in \R^d$, number of stages $K$, sketch-size $[m_1,...,m_K]$, sketched forward operator $[A_{s_1},... A_{s_K}]$, sketching operators $[\Sk_1,...,\Sk_K]$, up-sampling operators $[\U_1,...,\U_K]$, number of inner-loops for each stage $[N_1,...,N_K]$, step-size sequence $[\eta_1,...,\eta_{\sum_{k=1}^K N_k}]$, $\alpha \in (0, 1]$,  iteration counter $i = 0$
   \State{{\bfseries For} $k = 1$ {\bfseries to} $K$}
   \State{$\ \ ${\bfseries For} $j =1$ {\bfseries to} $N_k$}
\State $\ \ \ \ \ i \leftarrow i + 1$
\State $\ \ \ \ $ Generate random subsampling mask $M_i$
 \State $\ \ \ \ $ Compute the image-domain sketch: $v = \Sk_k(x_i)$
   \State $\ \ \ \ \ $Compute gradient estimate $G := \triangledown_{v} f(M_ib, M_i A_{s_k} v)$ 
   \State        $\ \ \ \ \ $$z_{i+1} =  x_i - \eta_i \U_k G,$
   \State     $\ \ \ \ \ $$x_{i+1} =  (1-\alpha)z_{i+1} + \alpha \D(z_{i+1}),$

  \State $\ \ ${\bfseries Endfor}
  \State {\bfseries Endfor}
\State Output $x_{i+1}$
\end{algorithmic}
\end{algorithm} 

To explain the motivation and derivation of Algorithm 1, we start by illustrating here a concrete example where the data-fidelity is the least-squares. Noting that the PnP proximal gradient descent iteration can be written as:
\begin{equation}\label{pnp}
     x_{k+1} =  \D[x_k - \eta \cdot (A^TA x_k - A^T b)],
\end{equation}
where $\D(\cdot)$ denotes the denoiser, which can be a denoising algorithm such as NLM/BM3D/TNRD, or a classical proximal operator of some convex regularization (such as TV-prox), or a pretrained denoising deep network such as (DnCNN). Then our sketched gradient can be written as:
\begin{equation}
    x_{k+1} =  \D[x_k - \eta \cdot \U(A_s^TA_s \Sk(x_k) - A_s^Tb)],
\end{equation}
where $\U(\cdot)$ denotes the upsamling operator. Numerically we found that off-the-shelf up/down-sampling algorithms such as the bi-cubic interpolation suffice to provide us good estimates of the true gradients. Using this scheme, an efficient approximation of the true gradient can be obtained since $A_s$ only takes a fraction of the computation of $A$, and usually $\U$ and $\Sk$ can be very efficiently computed. 

To further reduce the computational complexity, we can also utilize stochastic gradient estimate:
\begin{equation}
    x_{k+1} =  \D[x_k - \eta_k \cdot \U((M_kA_s)^T M_kA_s \Sk(x_k) - (M_kA_s)^T M_kb)]
\end{equation}
where $M_k$ is uniformly sampled minibatch of $I_{n \times n}$ here for computing the stochastic gradient. Here we use a vanilla minibatch stochastic gradient estimator. We can also choose here those advanced stochastic variance-reduced gradient estimator \citep{2012_Roux_Stochastic,johnson2013accelerating,defazio2014saga,driggs2020spring} for potentially even faster convergence.

In Algorithm 1 we present our PnP-MS2G framework, where we typically start by an aggressive sketch $\{A_{s_1}, \Sk_{1}\}$ with $m_1 \ll d$ for very fast initial convergence, and then for later stages we switch to medium size sketches $\{A_{s_k}, \Sk_{k}\}$ with $m_k < d$ which are increasingly more conservative, to reach a similar reconstruction accuracy as the unsketched counterpart.

We also wish to point out that in our sketching framework both the denoiser $\D$, the upsampling function $\U$ and sketching function $\Sk$ can be parameterized as deep (convolutional) neural-networks and trained either in a recursive or end-to-end manner, resulting in a new efficient deep-unrolling scheme \citep{adler2018learned,tang2021stochastic}.

In the multi-sketch framework presented here we gradually increase the sketch size $m$ throughout stages.

\subsection{Stochastic Lazy Denoisers}

In many scenarios in imaging applications, the computational costs of the proximal operators/PnP denoisers are also significant comparing to the costs of evaluating the gradients of the data-fidelity. Recall that our scheme can be summarized in one line:
\begin{equation}
    x_{k+1} =  \D[x_k - \eta_k \cdot \U((M_kA_s)^T M_kA_s \Sk(x_k) - (M_kA_s)^T M_kb)].
\end{equation}
Here one can further improve the efficiency in the early iteration if we replace the full operator $\D$ with a sketched/down-scaled version $\D_s$, such that $\D(\cdot) = \U\circ\D_s \circ\Sk(\cdot)$, then we have:
\begin{equation}\label{sk_denoise}
    x_{k+1} =  \U\circ\D_s \circ\Sk[x_k - \eta_k \cdot \U((M_kA_s)^T M_kA_s \Sk(x_k) - (M_kA_s)^T M_kb)].
\end{equation}
With this sketching scheme which we perform triple dimensionality reduction (both on the forward operator and denoiser), we can  accelerate PnP algorithms in reducing the complexity on denoiser but only in early iterations. In this subsection we propose two much more powerful schemes for accelerating PnP's denoiser computation based on the finding that the denoiser computation can actually be skipped with high-probability in each iteration, and hence further accelerates our sketched gradient schemes for PnP.

\paragraph{Lazy-PnP} The computational overhead of computing the denoiser can be effectively reduced via avoiding computing the denoiser at each of the iteration of PnP. We propose Lazy-Denoiser framework alongside with sketching, inspired by the ProxSkip algorithm \citep{papoutsellis2024we} used for convex optimization and federated learning \citep{mishchenko2022proxskip}. We present our Lazy-PnP scheme in Alg. \ref{A_lazy}, which allows us to execute the denoiser in only a fraction of iterations, while maintaining the same convergence rates and reconstruction accuracy in practice.

The Lazy-PnP scheme presented in Alg. \ref{A_lazy} utilize an auxiliary variable $h$ throughout the iterations for stabilization. This scheme is a stochastic approach which calls the denoiser with a relatively small probability $p$ (in our experiments we choose $p=20\%$ and $p=50\%$), with the denoising step written as \[x_{i+1} = \D(z_{i+1} - \frac{\eta}{p} h_i)\], otherwise the denoising step is skipped ($x_{i+1} = z_{i+1}$). If the denoiser step is execute in one iteration, then the auxiliary variable $h_i$ is also updated \[h_{i+1} = h_i + \frac{p}{\eta}(x_{i+1} - z_{i+1}).\] While performing the gradient descent step, the auxiliary variable $h$ is included \[z_{i+1} =  x_i - \eta (G - h_i).\] Since the variable $h_i$ keeps the average of implicit gradients of the denoiser, it can successfully compensate the fact that for the most of the iterations the denoiser is skipped, while keeping the algorithm numerically stable. Numerically we observe that we can easily skip $50\% - 80\%$ of the denoising steps while maintainly the same convergence rates for gradient-based PnP algorithms.

\begin{algorithm}[t]
   \caption{---Lazy PnP}
\begin{algorithmic}

   \State  {\bfseries Initialization:} $x_0 \in \R^d$, $h_0 \in \R^d$, probability $p\in (0,1]$ for calling the denoiser at each iteration. For accelerating PnP-MS2G, one can replace its inner-loop with this algorithm.
   \State{\textbf{For} $i = 1$ {\bfseries to} $K$}
\State $\ \ \ \ $ Generate random subsampling mask $M_i$
   \State $\ \ \ \ \ $Compute gradient estimate $G := \triangledown_{v} f(M_ib, M_i A v)$ 
   \State        $\ \ \ \ \ $Compute: $z_{i+1} =  x_i - \eta (G - h_i),$
   \State     $\ \ \ \ \ $With probability $p$ execute: $x_{i+1} = \D(z_{i+1} - \frac{\eta}{p} h_i)$, otherwise $x_{i+1} = z_{i+1}$
   \State  $\ \ \ \ \ $Compute: $h_{i+1} = h_i + \frac{p}{\eta}(x_{i+1} - z_{i+1})$

   \State \textbf{Endfor}
\State Output $x_{i+1}$
\end{algorithmic}
\label{A_lazy}
\end{algorithm}

\begin{algorithm}[t]
   \caption{---Lazy PnP-EQ (with Equivariant Denoiser)}
\begin{algorithmic}

   \State  {\bfseries Initialization:} $x_0 \in \R^d$, $h_0 \in \R^d$, probability $p\in (0,1]$ for calling the denoiser at each iteration. For accelerating PnP-MS2G, one can replace its inner-loop with this algorithm. 
   \State{\textbf{For} $i = 1$ {\bfseries to} $K$}
\State $\ \ \ \ $ Generate random subsampling mask $M_i$ and group action $T_{g_i}$ where $g_i \sim \mathcal{G}$
   \State $\ \ \ \ \ $Compute gradient estimate $G := \triangledown_{v} f(M_ib, M_i A v)$ 
   \State        $\ \ \ \ \ $Compute: $z_{i+1} =  x_i - \eta (G - h_i),$
   \State     $\ \ \ \ \ $With probability $p$ execute: $x_{i+1} = T_{g_i}^{-1}\D(T_{g_i}(z_{i+1} - \frac{\eta}{p} h_i))$, otherwise $x_{i+1} = z_{i+1}$
   \State  $\ \ \ \ \ $Compute: $h_{i+1} = h_i + \frac{p}{\eta}(x_{i+1} - z_{i+1})$

   \State \textbf{Endfor}
\State Output $x_{i+1}$
\end{algorithmic}
\label{A_lazy_e}
\end{algorithm} 

\paragraph{Equivariant Lazy PnP}

Recently a simple way of boosting the performance and stability of PnP methods has been proposed, namely the equivariant PnP \citep{terris2024equivariant}. Suppose we denote unitary matrix $\{T_g\}_{g \in \mathcal{G}}$ as transforms for some group $\mathcal{G}$, the equivariant denoiser can be expressed as:
\begin{equation}
    \D_{\mathcal{G}}(x) = T_g^{-1} \D(T_gx), \ \ g \sim \mathcal{G}
\end{equation}
Typical choices of the transforms include rotations, translations and reflections, etc. This scheme was in fact first found numerically in the work of Zhang et al \citep{zhang2021plug}. In the work of Terris et al \citep{terris2024equivariant}, this approach is formally analyzed and studied, demonstrating remarkable performance in stabilizing the iterations of PnP algorithms with performance gains. As this scheme is crucial for the application of deep denoiser in PnP schemes, we also leverge this in our Lazy-PnP framework, leading to a new algorithm as a side contribution.

In our Algorithm \ref{A_lazy} and \ref{A_lazy_e} we present Lazy-PnP schemes with stochastic gradients only because of ease of reading. Note that the same technique can be easily merged fully with PnP-MS2G framework, we omit this to avoid redundant presentations.

\subsection{Accelerating the Deep Unrolling Schemes via Operator Sketching}

In this section, we will present our two schemes for accelerating deep unrolling networks.
\subsubsection{One side sketching: Subset Approximation} In our new approach, we propose to replace the forward and adjoint operators in the full-batch LPD network of \citep{adler2018learned}, with only subsets of it. The proposed network can be view as an unrolled version of stochastic PDHG \citep{chambolle2018stochastic} (but with ordered-subsets, and without variance-reduction). We partition the forward and adjoint operators into $m$ subsets, and also the corresponding measurement data. In each layer, we use only one of the subsets, in a cycling order. Let ${\bf{M}} := [M_0, M_1, M_2,..., M_{m-1}]$ be the set of subsampling operators, then the saddle-point problem (\ref{saddle}) can be rewritten as:
\begin{equation}\label{saddle1}
    [x^\star, y^\star] = \min_{x} \max_{y} \{r(x) + \sum_{i = 0}^{m-1}\langle M_iAx, y_i \rangle - f_{b_i}^*(y_i)\}.
\end{equation}
Utilizing this finite-sum structure, our learned stochastic primal-dual (LSPD) network can be described as\footnote{Alternatively, one may also consider an optional learned momentum acceleration by keeping the memory of the outputs of a number of previous layers: $x_{k+1} = \Pm_{\te_p^k}(X_k, \tau_k, (M_iA)^T y_{k+1})$ where $X_k = [x_k, x_{k-1},.. x_{k-M}]$, at the cost of extra computation and memory. For such case the input channel of the subnets would be $M+1$.}:
 \begin{eqnarray*}
 && \mathrm{\textbf{Learned Stochastic Primal-Dual (LSPD)}} \\&&- \mathrm{Initialize}\ x_0 \in \mb{R}^d \ y_0 \in \mb{R}^{n/m}\\
 &&\mathrm{For} \ \ \ k = 0, 1, 2,...,  K-1\\
&&\left\lfloor
\begin{array}{l}
i = \mod(k,m); \\\mathrm{(or\ pick\ } i\mathrm{\ from\ } [0, m-1]\mathrm{\ uniformly\ at\ random)}\\
y_{k+1} = \D_{\te_d^k} (y_k, \sigma_k, (M_iA) {x}_k , M_ib);\\
x_{k+1} = \Pm_{\te_p^k}(x_k, \tau_k, (M_iA)^T y_{k+1});\\
\end{array}
\right.
 \end{eqnarray*}
In the scenarios where the forward operator dominates the computation in the unrolling network, for the same number of layers, our LSPD network is approximately $m$-time more efficient than the full-batch LPD network in terms of computational complexity. The LSPD we presented here describes a framework of deep learning based methods depending the parameterization of the primal and dual subnetworks and how they are trained. In practice the LPD and LSPD networks usually achieve best performance when trained completely end-to-end. While being the most recommended in practice, when trained end-to-end, it is almost impossible to provide any non-trivial theoretical guarantees. An alternative approach is to restrict the subnetworks across layers to be the same and train the subnetwork to perform denoising \citep{kamilov2017plug, sun2019online,tang2020fast, ono2017primal}, artifact removal \citep{liu2020rare}, or approximate projection to a image manifold \citep{rick2017one}, leading to a plug-and-play \citep{venkatakrishnan2013plug, romano2017little, reehorst2018regularization} type of approach with theoretical convergence guarantees.

 \begin{figure}[t]
   \centering

    {\includegraphics[width= .75\textwidth]{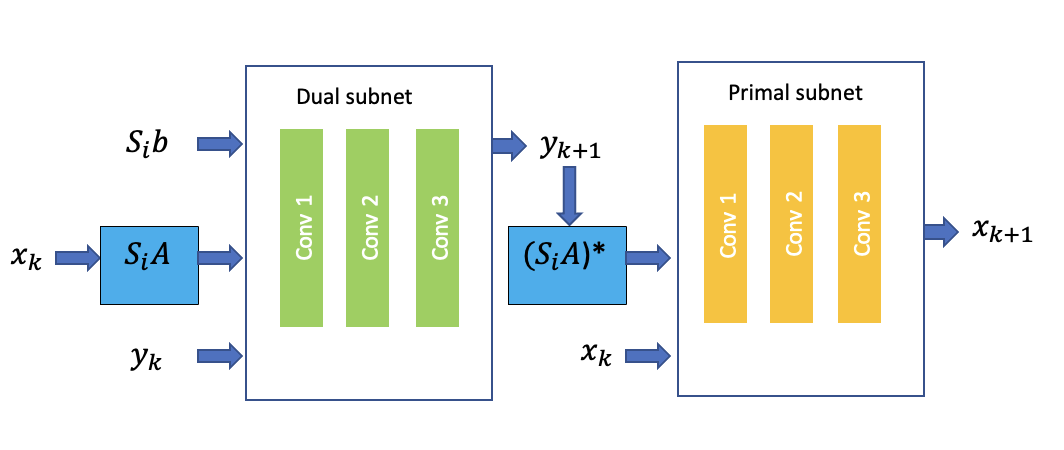}}
    \caption{One simple example of the practical choices for the building blocks of one layer of our LSPD network. Both dual and primal subnetworks are consist of 3 convolutional layers. The dual subnet has 3 input channels concatenating $[M_ib, M_iAx_k, y_k]$, while the primal subnet has 2 input channels for $[(M_iA)^Ty_{k+1}, x_k]$}

\end{figure} 

Note that our LSPD network covers the SGD-Net of \citep{liu2021sgd} as a special case, by setting the dual network to be a simple subtraction, and limit the primal network to only have one input channel taking in the stochastic gradient descent step with a fixed primal scalar step-size. We refer to this type of networks as the Learned SGD (LSGD) in this paper:
 \begin{eqnarray*}
 && \mathrm{\textbf{LSGD}} - \mathrm{Initialize}\ x_0 \in \mb{R}^d \ y_0 \in \mb{R}^{n/m}\\
 &&\mathrm{For} \ \ \ k = 0, 1, 2,...,  K-1\\
&&\left\lfloor
\begin{array}{l}
\mathrm{Pick\ } i\mathrm{\ from\ } [0, m-1]\mathrm{\ uniformly\ at\ random}\\
y_{k+1} =  M_iA{x}_k - M_ib;\\
x_{k+1} = \Pm_{\te_p^k}(x_k - \tau \cdot (M_iA)^T y_{k+1});\\
\end{array}
\right.
 \end{eqnarray*}
which is a stochastic variant of the ISTA-Net \citep{zhang2018ista}. The stochastic unrolling
can potentially give $m$ fold reduction in the per-iteration
complexity of unrolling.
 
\subsubsection{Double-sided Operator Sketching}
 Now we are ready to present our sketched LPD and sketched LSPD networks. Our main idea is to speedily approximate the products $A {x}_k$, $A^T y_{k+1}$:
 \begin{equation}\label{skaf}
     A {x}_k \approx A_{s_k} \Sk_{\te_s^k}({x}_k), \ \ A^T y_{k+1} \approx \U_{\te_u^k}(A_{s_k}^T y_{k+1})
 \end{equation}
  where $\Sk_{\te_s^k}(\cdot): \R^d \rightarrow \R^{d_{s_k}}$ ($d_{s_k} < d$) being the sketching/downsampling operator which can be potentially trainable w.r.t parameters $\te_s^k$, while $A_{s_k} \in \R^{n \times d_{s_k}}$ is the sketched forward operator discretized on the reduced low-dimensional image space, and for the dual step we have $\U_{\te_u^k} : \R^{d_{s_k}} \rightarrow \R^d$ the upsampling operator which can also be trained. In practice, we found that it is actually suffice for us to just use the most simple off-the-shelf up/down-sampling operators in Pytorch for example the bilinear interpolation to deliver excellent performance for the sketched unrolling networks. Our Sketched LPD network is written as:
\begin{eqnarray*}
 && \mathrm{\textbf{Sketched LPD}} - \mathrm{Initialize}\ x_0 \in \mb{R}^d \ y_0 \in \mb{R}^n\\
 &&\mathrm{For} \ \ \ k = 0, 1, 2,...,  K-1\\
&&\left\lfloor
\begin{array}{l}
y_{k+1} = \D_{\te_d^k} (y_k, \sigma_k, A_{s_k} \Sk_{\te_s^k}({x}_k) , b);\\
x_{k+1} = \Pm_{\te_p^k}(x_k, \tau_k, \U_{\te_u^k}(A_{s_k}^T y_{k+1}));\\
\end{array}
\right.
 \end{eqnarray*}
Again, we can use the same approximation for stochastic gradient steps:
 \begin{equation}\label{ska}
 \begin{aligned}
     &(M_i A) {x}_k \approx (M_iA_{s_k}) \Sk_{\te_s^k}({x}_k), \\& (M_i A)^T y_{k+1} \approx \U_{\te_u^k}((M_iA_{s_k})^T y_{k+1}),
 \end{aligned}
 \end{equation}
and hence we can write our Sketched LSPD (SkLSPD) network as:
  \begin{eqnarray*}
 && \mathrm{\textbf{SkLSPD} (Option 1)} \\&& - \mathrm{Initialize}\ x_0 \in \mb{R}^d \ y_0 \in \mb{R}^{n/m}\\
 &&\mathrm{For} \ \ \ k = 0, 1, 2,...,  K-1\\
&&\left\lfloor
\begin{array}{l}
i = \mod(k,m); \\\mathrm{(or\ pick\ } i\mathrm{\ from\ } [0, m-1]\mathrm{\ uniformly\ at\ random)}\\
y_{k+1} = \D_{\te_d^k} (y_k, \sigma_k, (M_iA_{s_k}) \Sk_{\te_s^k}({x}_k) , M_ib);\\
x_{k+1} = \Pm_{\te_p^k}(x_k, \tau_k, \U_{\te_u^k}((M_iA_{s_k})^T y_{k+1}));\\
\end{array}
\right.
 \end{eqnarray*}
 
or alternatively:
  \begin{eqnarray*}
 && \mathrm{\textbf{SkLSPD} (Option 2)} \\&&- \mathrm{Initialize}\ x_0 \in \mb{R}^d \ y_0 \in \mb{R}^{n/m}\\
 &&\mathrm{For} \ \ \ k = 0, 1, 2,...,  K-1\\
&&\left\lfloor
\begin{array}{l}
i = \mod(k,m); \\\mathrm{(or\ pick\ } i\mathrm{\ from\ } [0, m-1]\mathrm{\ uniformly\ at\ random)}\\
y_{k+1} = \D_{\te_d^k} (y_k, \sigma_k, (M_iA_{s_k}) \Sk_{\te_s^k}({x}_k) , M_ib);\\
x_{k+1} = \U_{\te_u^k}(\Pm_{\te_p^k}(\Sk_{\te_s^k}(x_k), \tau_k, (M_iA_{s_k})^T y_{k+1}));\\
\end{array}
\right.
 \end{eqnarray*}
 \paragraph{Remark regarding varying ``coarse-to-fine'' sketch size for SkLPD and SkLSPD} Numerically we suggest that we should use more aggressive sketch at the beginning for efficiency, while conservative sketch or non-sketch at latter iterations for accuracy. One plausible choice we found numerically pretty successful is: for the last few unrolling layers of SkLPD and SkLSPD, we switch to usual LPD/LSPD (say if the number of unrolling layers is 20, we can choose last 4 unrolling layers to be unsketched, that is, $A_{s_k} = A$ for $k>K_{\mathrm{switch}}$), and we found numerically that the reconstruction accuracy is best preserved if we implement this scheme.

 \paragraph{Remark regarding the Option 2 for further improving efficiency} The second option of our SkLPD and SkLSPD further accelerates the computation comparing to Option 1, by making the primal-subnet taking the low-dimensional images and gradients as input and then upscale.  Noting that the usual choice for the up and down sampler would simply be an off-the-shelf interpolation algorithm such as bilinear or bicubic interpolation which can be very efficiently computed, in practice we found the optional 2 often more favorable computationally if we use the coarse-to-fine sketch size. Numerically we found SkLPD and SkLSPD with option 2 and coarse-to-fine sketch size can be both trained faster and more efficient in testing due to the further reduction in the computation of the primal-subnet, without loss on reconstruction accuracy comparing to option 1.

\section{Theoretical Analysis}

In this section we provide theoretical recovery analysis of our operator sketching framework presented in the previous section. From this motivational analysis, we aim to demonstrate the reconstruction guarantee of PnP-MS2G and compare it with the recovery guarantee of the PnP-PGD/PnP-SGD derived under the same setting.

\subsection{General Assumptions}
We list here the assumptions we make in our motivational analysis of our generic sketching framework:
\begin{equation}
    x_{k+1} =  \Pm_{\te}[x_k - \eta_k \cdot \U((M_kA_s)^T M_kA_s \Sk(x_k) - (M_kA_s)^T M_kb)]
\end{equation}
\begin{myh}
\textbf{(Approximate projection)} We assume that the denoiser is a $\ve$-approximate projection towards a manifold $\M$: 
\begin{equation}
    \Pm_{\te}(x) = e(x) + P_\M(x),
\end{equation}
where:
\begin{equation}
    P_\M(x) := \arg \min_{z \in \M} \|x - z\|_2^2,
\end{equation}
and,
\begin{equation}
    \|e(x)\|_2 \leq \ve_0 , \ \ \forall x \in \mb{R}^d.
\end{equation}
\end{myh}
Here we model the denoiser to be a $\ve$-projection towards a manifold. Note that in practice the image manifold $\M$ typically form a non-convex subset. We also make conditions on the image manifold as the following:
\begin{myh}
\textbf{(Interpolation)}We assume the ground-truth image $x^\dagger \in \M$, where $\M$ is a closed set.
\end{myh}
With this condition on the manifold, we further assume restricted eigenvalue (restricted strong-convexity) condition which is necessary for robust recovery \citep{negahban2012unified,agarwal2012fast,wainwright2014structured,oymak2017sharp}:
\begin{myh}
\textbf{(Restricted Eigenvalue Condition)} We define a descent cone $\mathcal{C}$ at point $x^\dagger$ as:
 \begin{equation}
    \mathcal{C} := \left\{v \in \mathbb{R}^d |\  v = a(x - x^\dagger) , \forall a \geq 0, x \in \mathcal{M} \right\},
\end{equation}
and the restricted strong-convexity constant $\mathcal{\mu}_c$ to be the largest positive constant satisfies the following: 
 \begin{equation}
    \frac{1}{n} \|Av\|_2^2 \geq \mu_c \|v\|_2^2 ,\ \ \forall v \in \mathcal{C}.
 \end{equation}
 and the restricted smoothness constant $L_c$ to be the smallest positive constant satisfies:
 \begin{equation}
    \frac{1}{q} \|M_iAv\|_2^2 \leq L_c \|v\|_2^2 ,\ \ \forall v \in \mathcal{C}. \ \forall i \in [m]
 \end{equation} 
\end{myh}
The restricted eigenvalue condition is standard and crucial for robust estimation guarantee for linear inverse problems, i.e., for a linear inverse problem to be non-degenerate, such type of condition must hold \citep{oymak2017sharp, agarwal2012fast, 2012_Chandrasekaran_Convex}. For example, in sparse-recovery setting, when the measurement operator is a Gaussian map (compressed-sening measurements) and $x^\dagger$ is $s$-sparse, one can show that $\mu_c$ can be as large as $O(1 - \frac{s \log d}{n})$ \citep{oymak2017sharp}. In our setting we would expect an even better $\mu_c$, since the mainifold of certain classes of real-world images should have much smaller covering numbers compared to the sparse set.

\subsection{Estimation error bounds of PnP-MS2G}
With the assumptions presented in the previous subsection, here we provide the recovery guarantee of PnP-MS2G on linear inverse problem where we have $b = Ax^\dagger$. Denoting $L_s$ to be the smallest constant satisfying:
\begin{equation}
    \frac{1}{q}\|M_iAv\|_2^2 \leq L_s \|v\|_2^2, \ \ \forall v \in \mb{R}^d, i \in [m],
\end{equation}
we can have the following result:
\begin{theorem}\label{thm1}
(Upper bound) Assuming \textbf{A.1-3}, let $\eta = \frac{1}{qL_s}$ and $b = Ax^\dagger + w$, the output of $k$-th stage of PnP-MS2G has the following guarantee for the estimation of $x^\dagger$:
\begin{equation}
    \E\|x_{N_k} - x^\dagger\|_2 \leq \alpha^{N_k} \|x_{\mr{init}} - x^\dagger\|_2+ \frac{1 - \alpha^{{N_k}}}{1 - \alpha} (\ve + \delta),
\end{equation}
where $x_{\mr{init}}$ denotes the initial point of $k$-th stage of PnP-MS2G, $\alpha = \kappa(1 - \frac{\mu_c}{ L_s})$, $\kappa=1$ if $\M$ is convex, $\kappa=2$ if $\M$ is non-convex, $\ve = \ve_0 + \kappa\eta\ve_1 + \kappa\eta\ve_2$ and let ,
\begin{equation}
\begin{split}
&\delta:= 2\eta\E\sup_{v \in \mathcal{C} \cap \mathcal{B}^{d}, i \in [m]} v^TA^TM_i^TM_iw\\
        &\|M_iA_{s_k}\|_2\|M_iA_{s_k} \D(x_k) - M_iAx_k\|_2 \leq \ve_1, \forall i, k \\
        &\|\U(M_iA_{s_k})^Ty_k - (M_iA)^Ty_k\|_2 \leq \ve_2, \forall i, k
\end{split}
\end{equation}

\end{theorem}


When the restricted eigenvalue $\mu_c$ is large enough such that $\alpha < 1$, the PnP-MS2G has a linear convergence in estimation error, up to $\frac{\ve}{1 - \alpha}$ only depending the approximation accuracy of the denoiser in terms of projection. For many inverse problems for example CT/PET tomographic imaging we have $L_s \approx L_f$ where $L_f$ being the largest eigenvalue of $\frac{1}{n}A^TA$, and in these tasks the same convergence rate in Theorem \ref{thm1} apply for both sketched algorithms and the full-batch counterpart. This suggests the tremendous potential of computational saving of PnP-MS2G over classical methods.

From the above bound we can observe that, the reconstruction accuracy of a certain stage of PnP-MS2G is dependent on $\ve_1$ and $\ve_2$ which directly rely on the sketch-size on the image domain. If we eventually reduce the image-domain sketch-size, then $\ve_1, \ve_2 \rightarrow 0$ and we can have optimal estimation accuracy. Hence this bound demonstrates that our multi-stage strategy is necessary.

One the other hand, using similar technique we can provide a complementing lower bound for the estimation error of PnP-MS2G:
\begin{theorem}
 (Lower bound.) Under the same conditions of Theorem \ref{thm1}, if we further assume the constraint set $\M$ is convex, for any $\gamma > 0$, $\exists R(\gamma)$, if $\|x_{\mr{init}} - x^\dagger\|_2 \leq R(\gamma)$, the estimation error of  the output of the $k$-th stage PnP-MS2G satisfies the lower bound:
 \begin{equation}
     \E\|x_{N_k} - x^\dagger\|_2 \geq (1 - \gamma)^{N_k} (1 - \frac{L_c}{L_s})^{N_k} \|x_{\mr{init}} - x^\dagger\|_2 - \frac{L_s}{L_c} \ve_\star
 \end{equation}
 where $\ve_\star = \ve_0 + \kappa\eta\ve_1 + \kappa\eta\ve_2$.
\end{theorem}
Again, we present the proof of this result in the appendix.

\begin{figure}[t]
  
    {\includegraphics[width=0.95\textwidth]{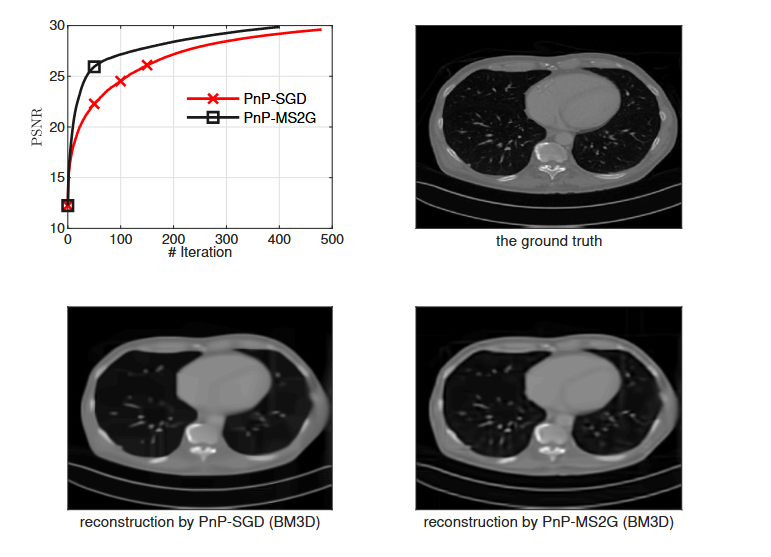}}
  
   \caption{Example for applying MS2G (minibatched) in X-ray CT reconstruction, comparing to the PnP stochastic gradient descent (PnP-SGD) proposed by Sun et al \citep{sun2019online}. Note that each iteration of PnP-MS2G is more computationally efficient than PnP-SGD due to the dimensionality reduction by operator sketching. Surprisingly, even in terms of iteration-number the PnP-MS2G can provide better convergence rate comparing to PnP-SGD.}
    \label{f_m1}
\end{figure}

  
  

\begin{figure}[t]
  
    {\includegraphics[width=0.95\textwidth]{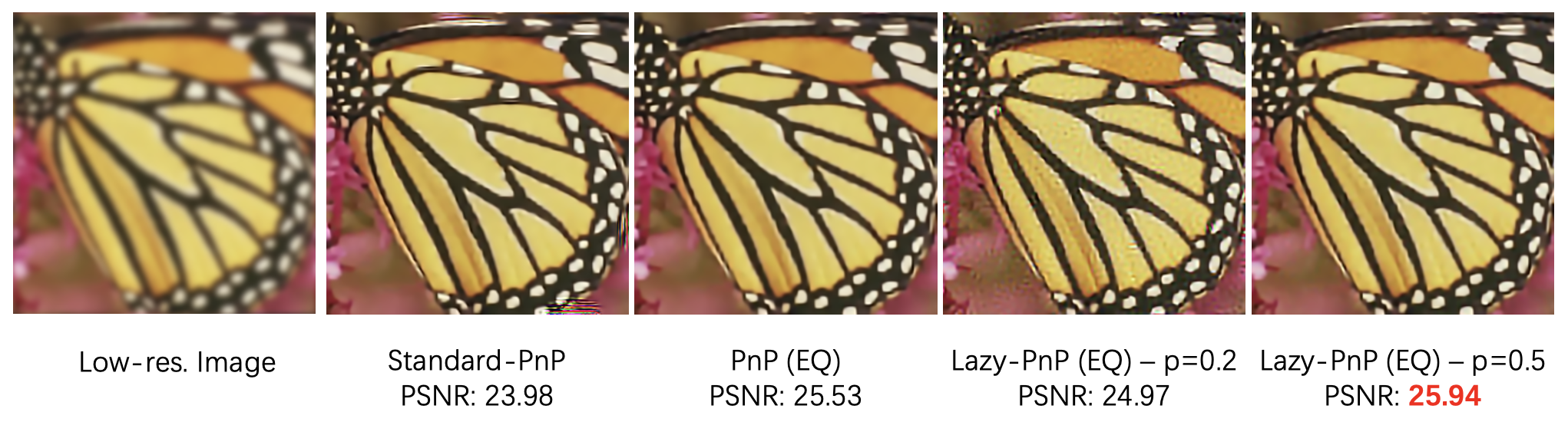}}
{\includegraphics[width=0.95\textwidth]{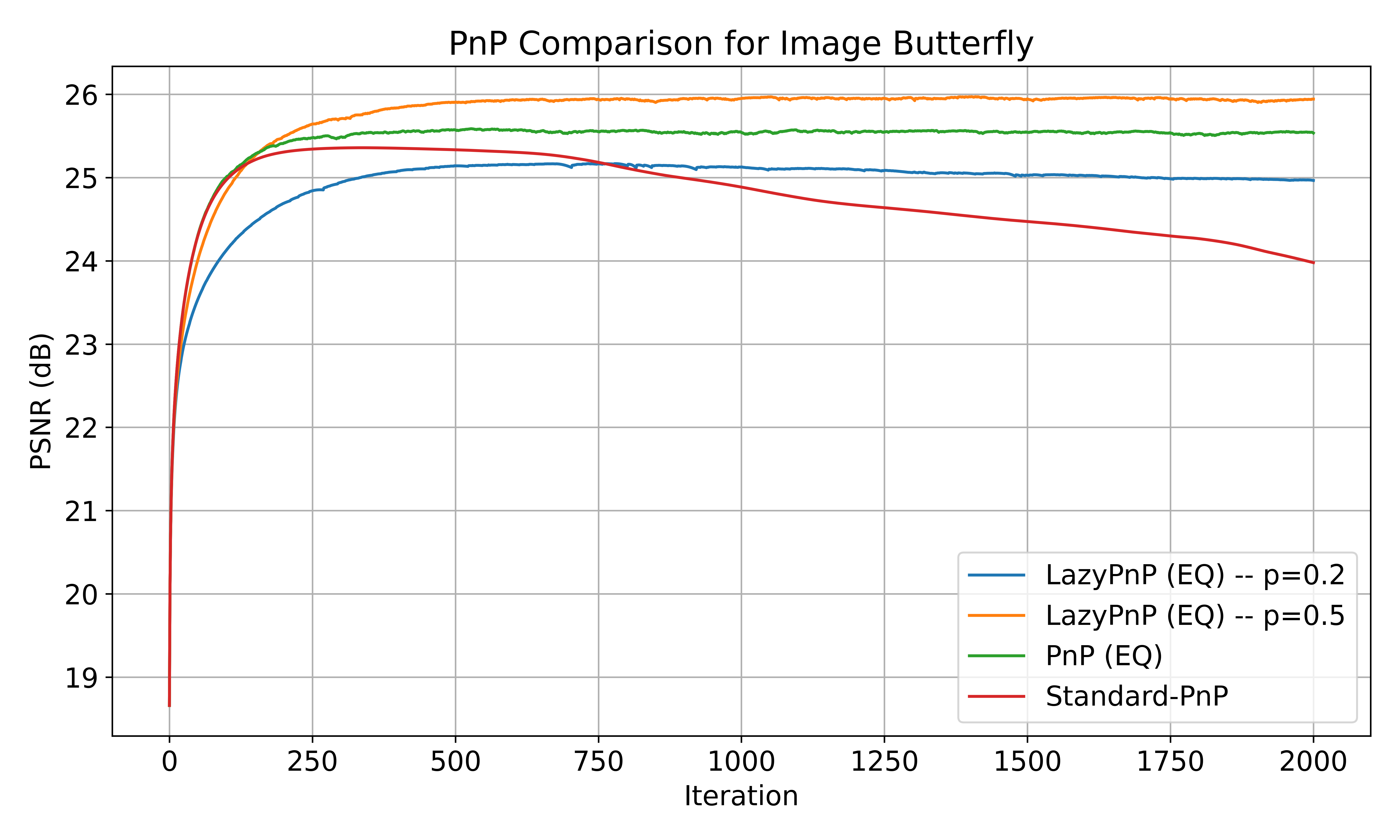}}
   
  
   \caption{\textbf{(Lazy-Denoiser)}Example for applying our Lazy-Denoiser scheme with equivariant-PnP for image superresolution ($4\times$). The denoiser network we choose here is DnCNN. Here we show the reconstructed image at 2000-th iteration.}
    \label{f_a51}
\end{figure}

\begin{figure}[t]
  
     {\includegraphics[width=0.95\textwidth]{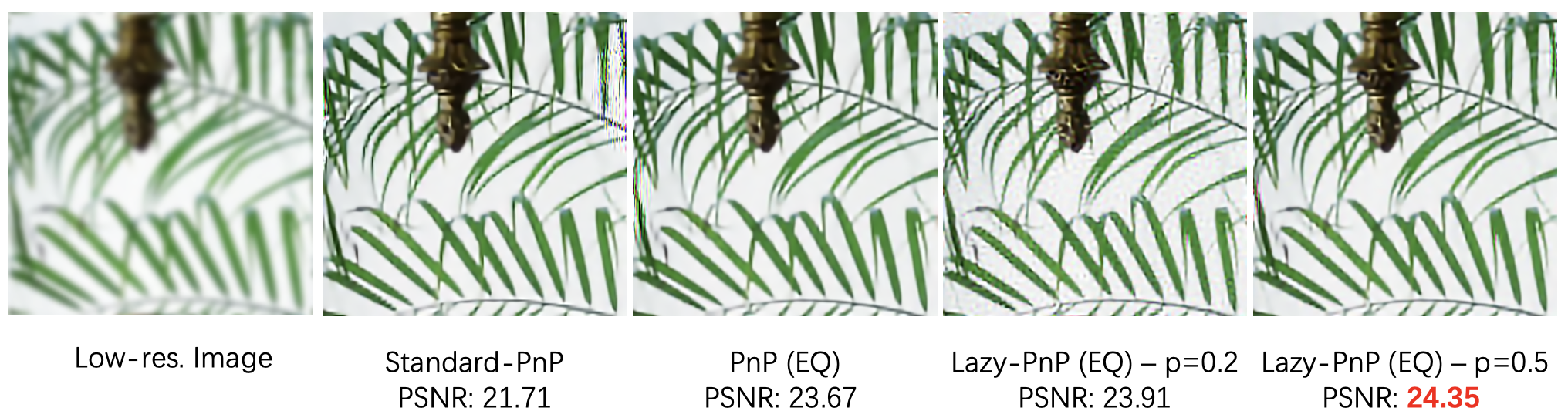}}
    {\includegraphics[width=0.95\textwidth]{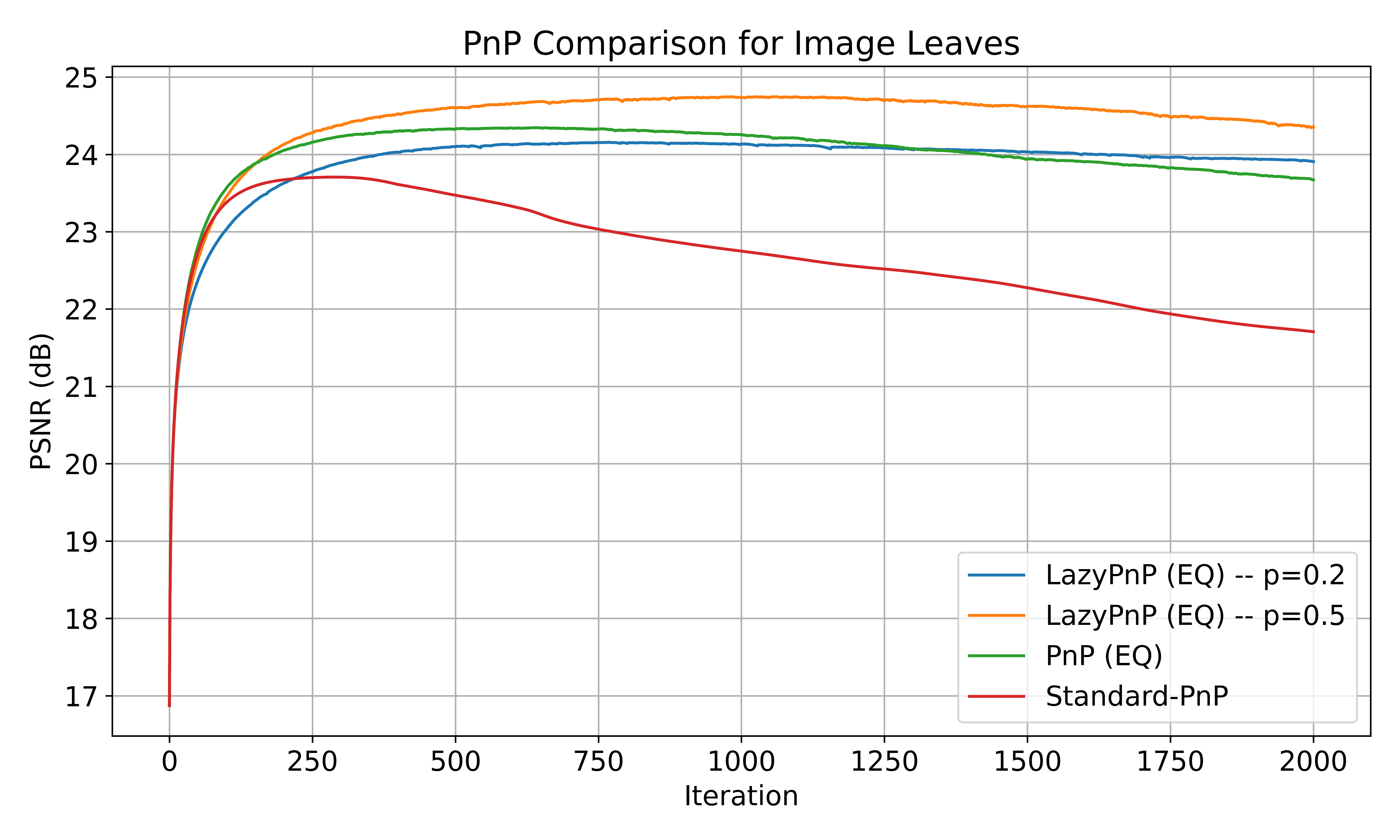}}  
   \caption{\textbf{(Lazy-Denoiser)}Example for applying our Lazy-Denoiser scheme with equivariant-PnP for image superresolution ($4\times$). The denoiser network we choose here is DnCNN. Here we show the reconstructed image at 2000-th iteration.}
    \label{f_a52}
\end{figure}

\begin{figure}[t]
  
     {\includegraphics[width=0.95\textwidth]{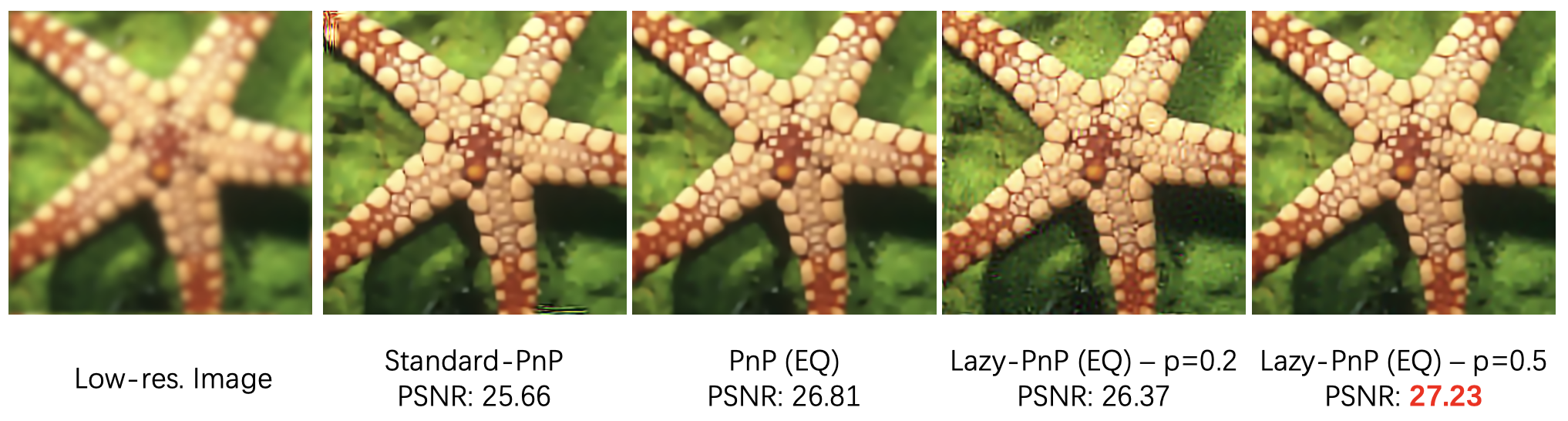}}
    {\includegraphics[width=0.95\textwidth]{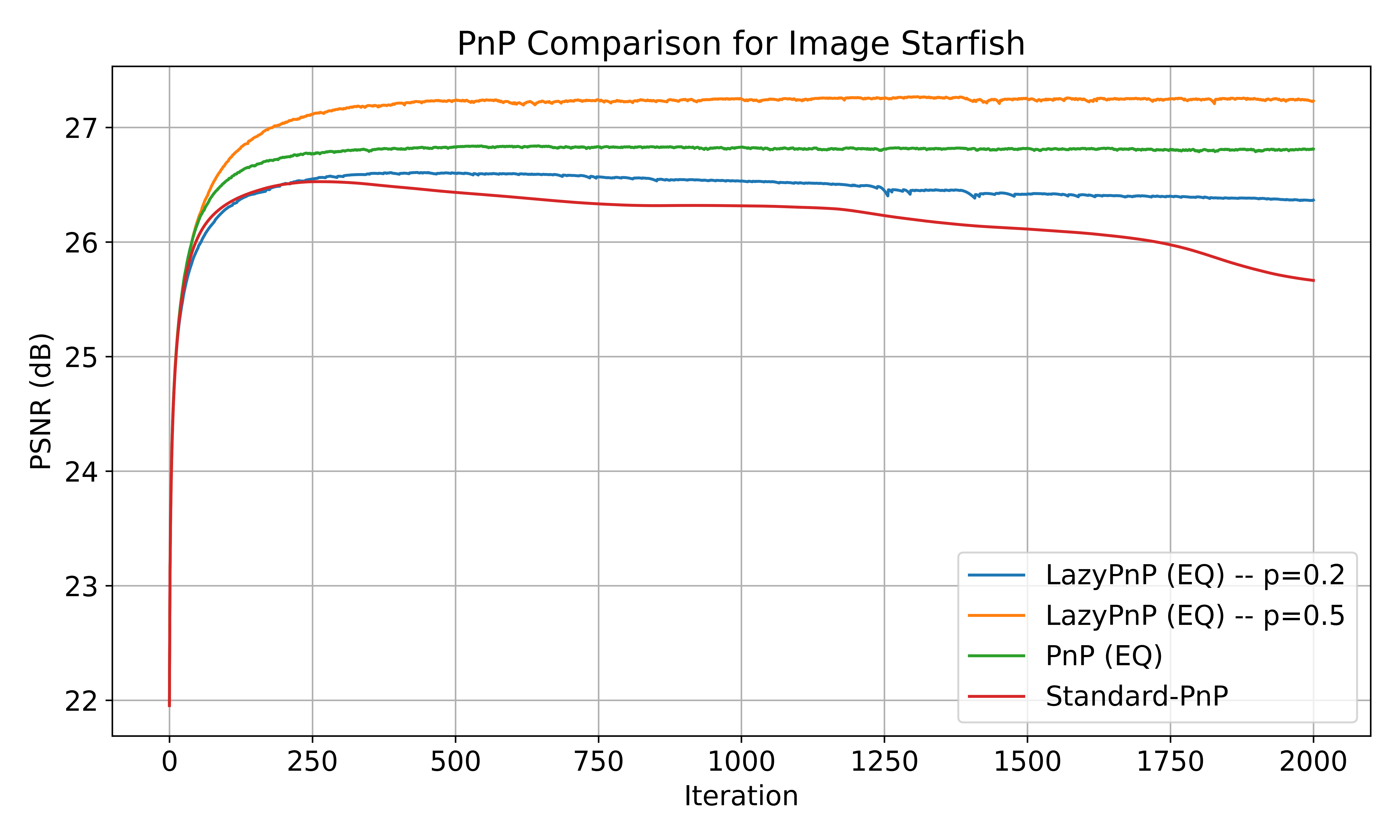}}  
   \caption{\textbf{(Lazy-Denoiser)}Example for applying our Lazy-Denoiser scheme with equivariant-PnP for image superresolution ($4\times$). The denoiser network we choose here is DnCNN. Here we show the reconstructed image at 2000-th iteration.}
    \label{f_a53}
\end{figure}

\begin{figure}[t]
  
    {\includegraphics[width=0.95\textwidth]{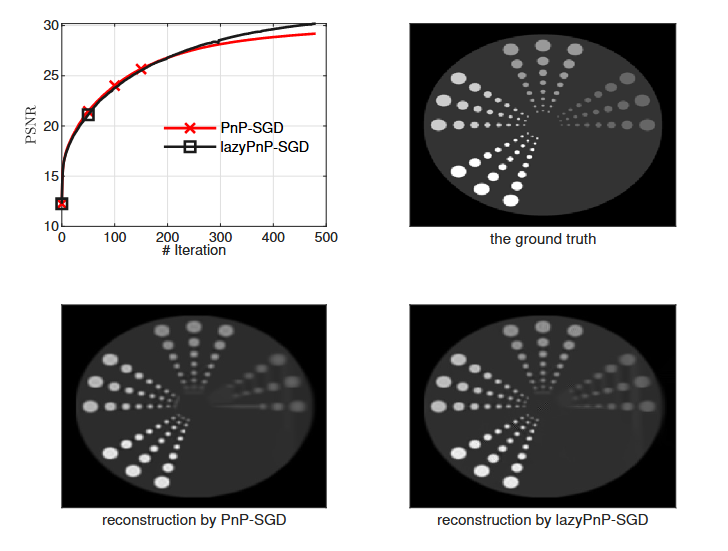}}
  
   \caption{\textbf{(Lazy-Denoiser)}Example for applying our Lazy-Denoiser scheme in X-ray CT reconstruction. We choose $p= \frac{1}{5}$ for our Lazy-PnP-SGD, which means that the number of calls on the denoiser for our Lazy-PnP-SGD is only $20\%$ of the standard PnP-SGD, while maintaining the same convergence rate.}
    \label{f_a5}
\end{figure}

\section{Numerical Experiments}

\subsection{Numerical study for Sketched Plug-and-Play algorithms}

We start by presenting the numerical results for applying our sketching framework on accelerating PnP algorithms. We start our illustration by sparse-view CT reconstruction tasks. Here we compare our PnP-MS2G with PnP-PGD and its stochastic variant PnP-SGD \citep{sun2019online}. For our PnP-MS2G we perform a $4\times$ downscale at the first 50 iterations, then a $2\times$ downscale afterwards, leading to significant improvment on computational efficiency. We choose here the famous BM3D \citep{dabov2007image,dabov2008image} denoiser. In the numerical results we reported in the figures, we found that surprisingly even in terms of the iteration counts our scheme can achieve an improvement on the convergence speed, let alone our PnP-MS2G is more efficient per iteration comparing to PnP-PGD and PnP-SGD.

In Figure \ref{f_a51},\ref{f_a52}, and \ref{f_a5} we present numerical results for our Lazy-Denoiser scheme on image superresolution and X-ray CT image reconstruction tasks. We first test our Lazy-PnP scheme with equivariant denoiser presented in Algorithm 3, comparing to standard equivariant PnP scheme \citep{terris2024equivariant} in image superresolution task. Here we seek to perform $4\times$ superresolution for low-resolution color images. The setting of this experiment is the same as the experiment in \citep{terris2024equivariant}, with The interpolation kernel $h$ for the task being the Guassian kernel of standard deviation 1:
\begin{equation}
    b = (h * x)_{\downarrow4} + \epsilon,
\end{equation}
where $\epsilon$ is a Gaussian noise with standard deviation 0.05. The denoiser we choose is the pretrained DnCNN. The numerical results for superresolution is demonstrated in Figure \ref{f_a51} and \ref{f_a52}, where we can observe that for this denoiser-dominant case, our Lazy-PnP (EQ) with $p=0.5$ consistently outperforms equivariant PnP scheme, which means we save $50\%$ of computation on the denoiser, while the standard PnP-PGD diverges.

In the X-ray CT experiment we presented in Figure \ref{f_a5}, we implement PnP-SGD with or without the Lazy-Denoiser scheme. We can observe that, our Lazy-PnP has the same convergence rates comparing to vanilla PnP-SGD, while only require to compute the denoiser on $20\%$ of the iteration.

\subsection{Training of LSPD and SkLSPD}
\subsubsection{Supervised end-to-end training}
The most basic training approach for LSPD/SkLSPD is the end-to-end supervised training where we consider fully paired training samples of measurement and the \say{ground-truth} -- which is typically obtained via a reconstruction from high-accuracy and abundant measurements. We take the initialization of LSPD/SkLSPD as a \say{filtered back-projection} $x^0 = A^\dagger b$. Let $\te$ be the set of parameters $\te := \{\te_p^k, \te_d^k, \tau_k, \sigma_k\}_{k=0}^{K-1}$, applying the LSPD/SkLSPD network on some measurement $b$ can be written as $\F_\te(b)$, the training objective can typically be written as:
\begin{equation}\label{sup_loss}
    \te^\star \in \arg\min_{\te} \sum_{i=1}^N \|x_i^\dagger - \F_{\te}(b_i, x^0_i)\|_2^2,
\end{equation}
where we denote by $N$ the number of paired training examples. Since the paired high-quality ground-truth data is expensive and idealistic in practice, we would prefer more advanced methods for training which have relaxed requirements on available training data. In Appendix, we present unsupervised training results for the unrolling networks using only measurement data.


In this subsection we present numerical results of our proposed networks for low-dose X-ray CT. In real world clinical practice, the low dosage CT is widely used and highly recommended, since the intense exposures to the X-ray could significantly increase the risk of inducing cancers. The low-dose CT takes a large number of low-energy X-ray views, leading to huge volumes of noisy measurements. This makes the reconstruction schemes struggle to achieve efficient and accurate estimations. In our X-ray CT experiments we use the standard Mayo-Clinic dataset \citep{mccollough2016tu} which contains 10 patients' 3D CT scans. We use 2111 slices (from 9 patients) of 2D images sized $512 \times 512$ for training and 118 slices of the remaining 1 patient for testing. We use the ODL toolbox \citep{adler2018learned} to simulate fan beam projection data with 800 equally-spaced angles of views (each view includes 800 rays). The fan-beam CT measurement is corrupted with Poisson noise: $b \sim \mr{Poisson}(I_0 e^{-Ax^\dagger})$, where we make a low-dose choice of $I_0=7 \times 10^4$.  We use the Beer-Lambert law to simulate the noisy projection data, and to linearize the measurements, we consider the log data.

In our LSPD and SkLSPD networks we interleave-partition (according to the angles) the forward/adjoint operators and data into $m=4$ subsets. Our networks has $K=12$ layers\footnote{each layer of LSPD includes a primal and a dual subnetwork with 3 convolutional layers with kernel size $5 \times 5$ and 32 channels, same for LPD.} hence correspond to 3 data-passes, which means it takes only 3 calls in total on the forward and adjoint operators. We compare it with the learned primal-dual (LPD) which has $K=12$ layers, corresponding to 12 calls on the forward and adjoint operator. We train all the networks with 50 epochs of Adam optimizer \citep{kingma2014adam} with batch size 1, in the supervised manner.

For our SkLSPD we choose the Option 2 presented in our Section III-B, with the coarse-to-fine sketch-size. For all these networks, we choose the subnetworks $\Pm_{\te_k}$ and $\D_{\te_k}$ to have 3 convolutional layers (with a skip connection between the first channel of input and the output) and 32 channels, with kernel size 5. The starting point $x_0$ is set to be the standard filtered-backprojection for all the unrolling networks. We set all of them to have 12 algorithmic layers ($K=12$). For the up/down-samplers in our Sketched LSPD, we simply choose the bilinear upsample and downsample functions in Pytorch. When called, the up-sampler increase the input image 4 times larger (from $256 \times 256$ to $512 \times 512$), while the down-sampler makes the input image 4 times smaller (from $512 \times 512$ to $256 \times 256$). While the full forward operator $A$ is defined on the grid of $512 \times 512$, the sketched operator $A_s$ is defined on the grid of $256 \times 256$ hence requires only a half of the computation in this setting. We use the coarse-to-fine strategy for SkLSPD, where we sketch the first 8 layers, but left the last 4 layers unsketched. We also implement and test the SkLSPD with a light-weight dual-subnetwork (corresponding to a proximal operator of a weighted $\ell_2$ loss with learnable weights, see the SkLSPD-LW in Section IV).

\begin{figure*}[t]
   \begin{center}
    {\includegraphics[trim=80 70 15
    35,clip,width=0.95\textwidth]{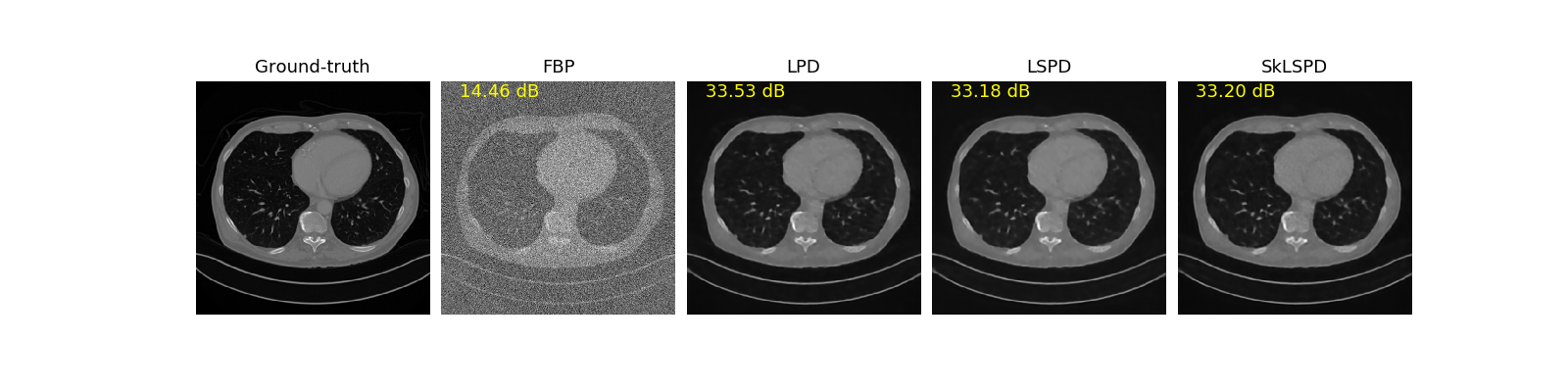}}
    {\includegraphics[trim=80 130 15 95,clip,width=0.95\textwidth]{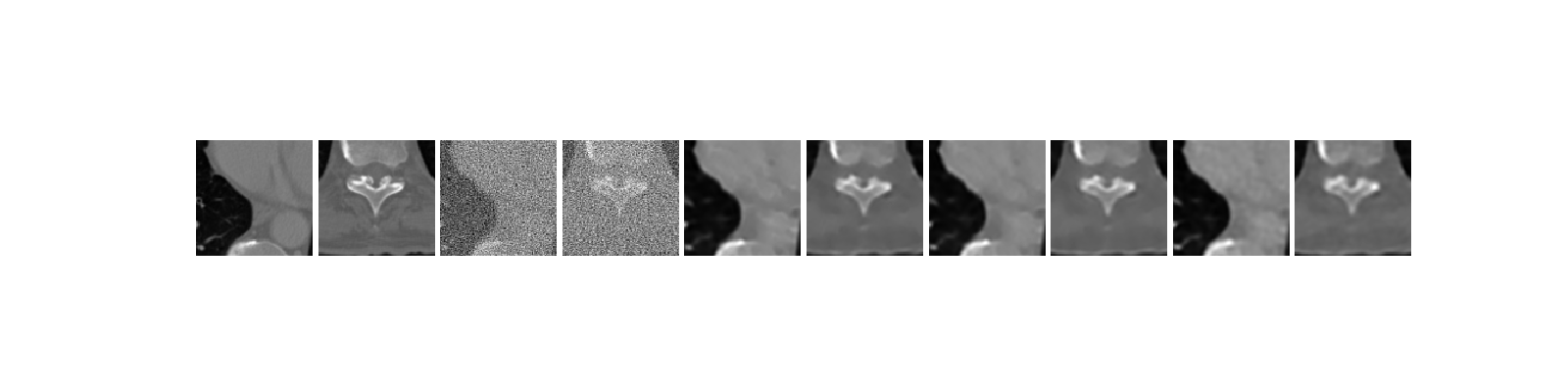}}
        {\includegraphics[trim=80 70 15 55,clip,width=0.95\textwidth]{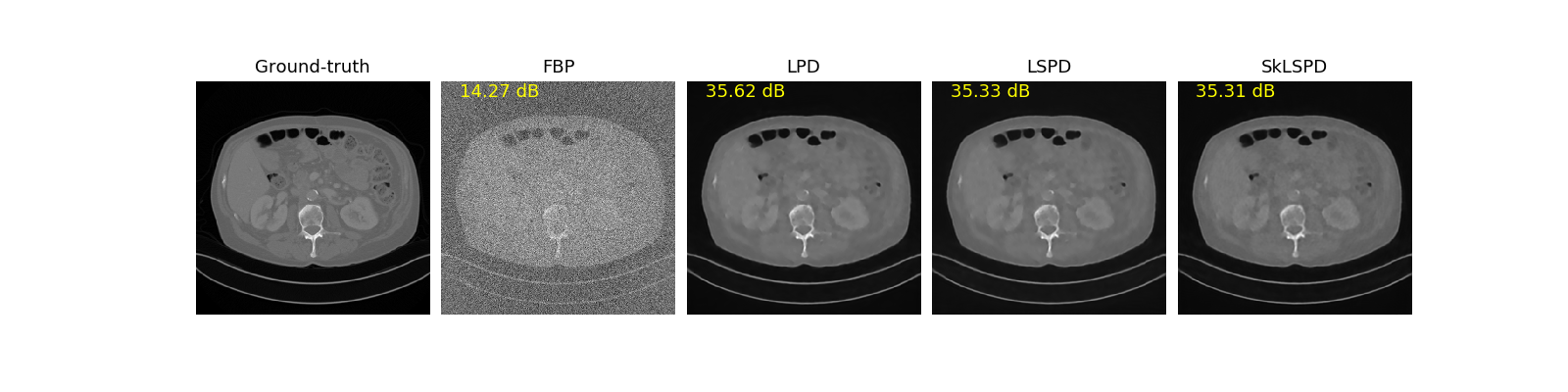}}
            {\includegraphics[trim=80 130 15 95,clip,width=0.95\textwidth]{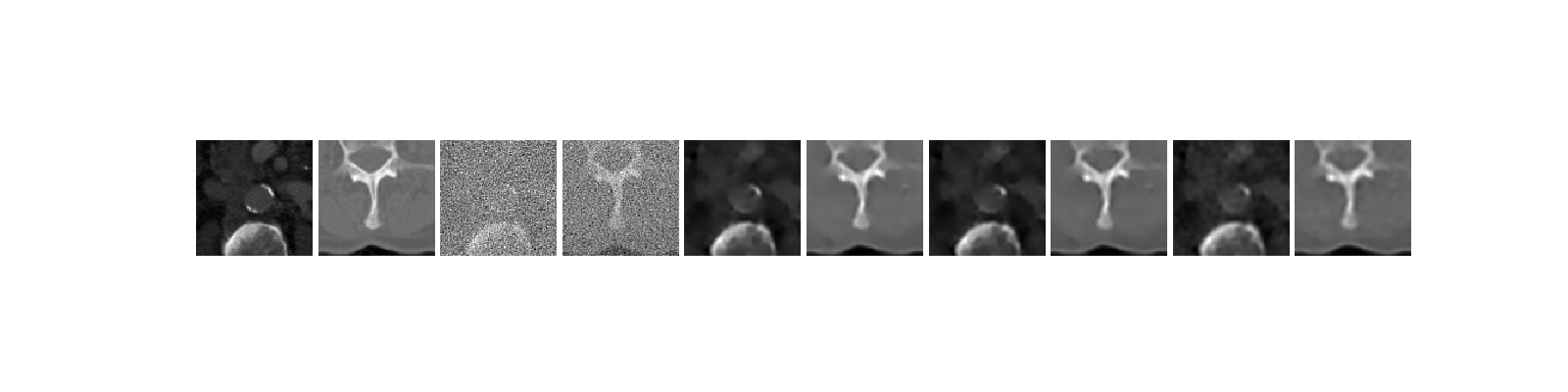}}
    \end{center}
   \caption{Examples for Low-dose CT on the test set of Mayo dataset. We can observe that our LSPD networks achieve the same reconstruction performance as the full-batch LPD}
    \label{f3}
\end{figure*}

\begin{figure}[t]
   \begin{center}
    {\includegraphics[trim=20 20 15 5,clip,width=0.95\textwidth]{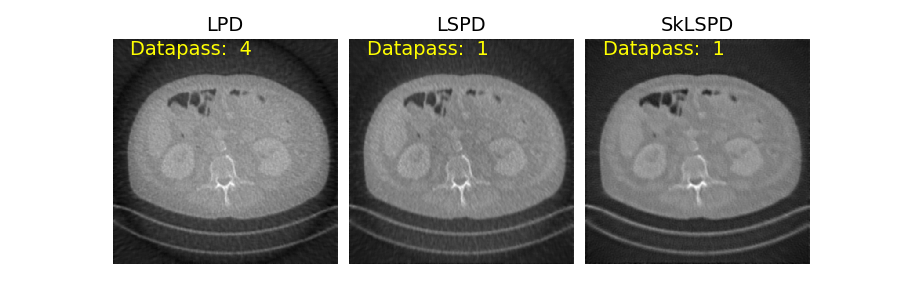}}
    {\includegraphics[trim=20 20 15 25,clip,width=0.95\textwidth]{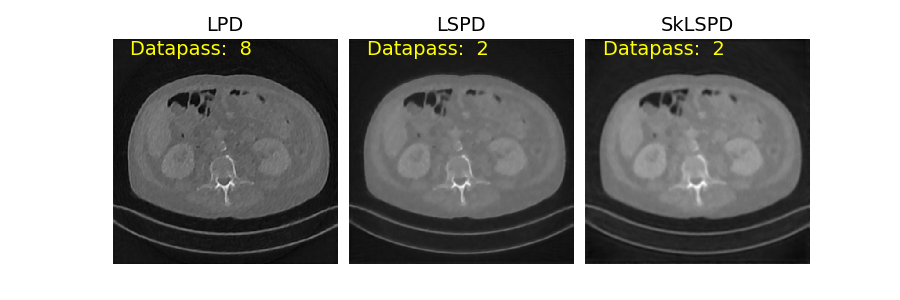}}
        {\includegraphics[trim=20 20 15
    25,clip,width=0.95\textwidth]{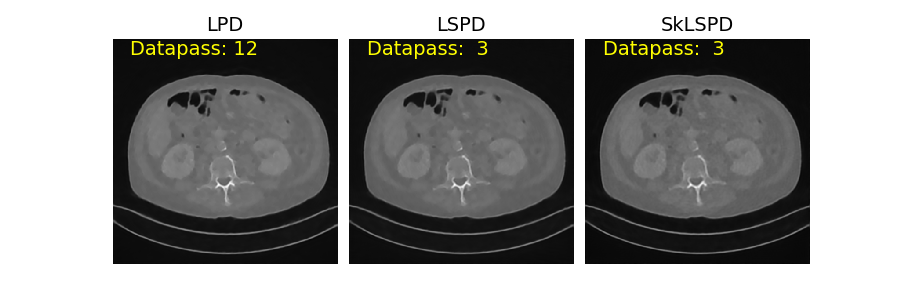}}
    \end{center}
   \caption{Example for intermediate layer outputs for Low-dose CT on the test set of Mayo dataset. We can observe that LSPD/SkLSPD achieves competitive reconstruction quality with LPD across intermediate layers.}
    \label{f7}
\end{figure}

\begin{table*}[t]\label{ld_table}
\caption{Low-dose CT testing results for LPD, LSPD and SkLSPD networks on Mayo dataset, with supervised training}
\label{sample-table}
\vskip 0.15in
\begin{center}
\begin{small}
\begin{sc}
\begin{tabular}{lcccr}
\hline
Method & $\#$ calls & PSNR & SSIM & GPU time (s) on\\ &on $A$ and $A^T$&&&  1 pass of
\\ &&&&  test set\\
\hline

&&&\\

FBP & - & 14.3242& 0.0663 &
\\
\\
LPD \textit{(12 layers)} & 24& 35.3177 & 0.9065 & 48.348 \\
LSGD \textit{(24 layers)} & 12& 31.5825& 0.8528& 33.089\\
LSPD  \textit{(12 layers)} & 6& 35.0577 & 0.9014 & 31.196\\
SkLSPD \textit{(12 layers)} & 4 & 34.9749 & 0.9028 & 23.996\\
SkLSPD \textit{(12 layers, } & 4 & 34.6389 & 0.8939 & 19.843\\
\textit{light weight on dual-step)} &&&&
\\
\hline
\end{tabular}
\end{sc}
\end{small}
\end{center}
\vskip -0.1in
\end{table*}

In addition, we also implement the Learned SGD \citep{liu2021sgd} in our setting which can be viewed as a simple special case of our LSPD network (see Section III-A) . Here for LSGD we choose the same parameterization of primal sub-networks as our LSPD (except by their original design the LSGD subnetworks only take 1 input channel). To make a fair comparison, since LSGD do not have dual-subnetworks, we allow the LSGD to have 24 layers, such that the total number learnable parameters is similar to our LSPD.

\begin{table*}[t]\label{sv_table}
\caption{Sparse-View CT testing results for LPD and SkLSPD networks on Mayo dataset, with supervised training}
\label{sample-table-1}
\vskip 0.15in
\begin{center}
\begin{small}
\begin{sc}
\begin{tabular}{lcccr}
\hline
Method & $\#$ calls & PSNR & SSIM & GPU time (s) on\\ &on $A$ and $A^T$&&&  1 pass of
\\ &&&&  test set\\
\hline

&&&\\

FBP & - & 22.0299& 0.2713 &
\\
\\
LPD \textit{(12 layers)} & 24& 36.9198 & 0.9129 & 28.018 \\
SkLSPD \textit{(12 layers)} & 4 & 36.6359 & 0.9178 & 17.340\\
\\
\hline
\end{tabular}
\end{sc}
\end{small}
\end{center}
\vskip -0.1in
\end{table*}

We present the performance of the LPD, LSPD, and SkLSPD on the test set in Table 1, and some illustrative examples in Figure 2 for a visual comparison. We also present the results of the classical Filtered-Backprojection (FBP) algorithm, which is widely used in clinical practice. We can observe from the FBP baseline, due to the challenging extreme low-dose setting, the FBP reconstruction fails completely. This can be partially addressed by U-Net postpocessing (FBPConvNet, \citep{jin2017deep}), whose parameter size is one order of magnitute larger than our unrolling networks. Next we turn to the learned reconstruction results. From the numerical results, we found out that our LSPD and SkLSPD networks both achieve almost the same reconstruction accuracy compare to LPD baseline in terms of PSNR (peak signal-to-noise ratio) and SSIM (structural similarity index, \citep{wang2004image}) measures, with requiring only a fraction of the computation of the forward and adjoint operators. In terms of run time on GPU, our acceleration can introduce a reduction of around $40\%$ to $60\%$ compared to the full batch LPD.

In our Table II, we present additional results on another widely applied modality in clinical practice -- the sparse-view CT, where we take fewer amount of normal-dose measurements. Here we use again the ODL toolbox to simulate fan beam projection data with 200 equally-spaced angles of views (each view includes 1200 rays). The fan-beam CT measurement is corrupted with Poisson noise: $b \sim \mr{Poisson}(I_0 e^{-Ax^\dagger})$, where we make a normal-dose choice of $I_0=7 \times 10^6$. Different to the low-dose CT, the main challenge of sparse-view CT is the ill-poseness of the inverse problems, that the measurement operator is highly under-determined with a non-trivial null-space. Meanwhile we also present in the Appendix our numerical results on instance-adaptation in out-of-distribution reconstruction for sparse-view CT, demonstrating again the effectiveness of our approach.

\section{Conclusion}

In this work we proposed a new paradigm for accelerating iterative data-driven reconstruction (IDR) schemes such as plug-and-play methods and deep unrolling networks, and we have done recovery analysis for such frameworks for the first time and perform a thorough comparison to full-batch unrolling. Our generic framework is based on leveraging the spirit of sketching in stochastic optimization and dimensionality reduction into the design of IDR schemes for computational efficiency and memory efficiency in solving large-scale imaging inverse problems. Moreover, we propose auxiliary denoiser sketching schemes for mitigating the computational overheads of advanced denoisers for plug-and-play methods. We have provided theoretical analysis of the proposed framework for the estimation guarantees from the viewpoint of stochastic optimization theory. Then we provide numerical study of the proposed schemes in the context of X-ray CT image reconstruction, demonstrating the effectiveness of our acceleration framework for deep unrolling networks. Although in this paper we mostly applied our sketching framework for accelerating PnP and deep unrolling, we need to emphasis here that this framework can be readily applied to accelerate newer algorithmic schemes such as deep restoration prior \citep{hu2023restoration} and deep equilibrium models \citep{gilton2021deep} and generative conditional sampling \citep{cai2024nf}.

\section{Appendix}

\subsection{Proof of Theorem 3.1}

In this proof we utilize several projection identities from \citep{oymak2017sharp}. We list them here first for completeness. The first one would be the cone-projection identity:
\begin{equation}
    \|P_\mathcal{C}(x)\|_2 = \sup_{v \in \mathcal{C} \cap \mathcal{B}^d} v^Tx,
\end{equation}
where $\mathcal{B}^d$ denotes the uni-ball in $\mb{R}^d$. The second one is the shift projection identity regarding that the Euclidean distance is preserved under translation:
\begin{equation}
    P_\M(x + v) - x = P_{\M - x} (v).
\end{equation}
Now if $0  \in \M - x$, we can have the third projection identity which is an important result from geometric functional analysis \citep[Lemma 18]{oymak2017sharp}:
\begin{equation}
    \|P_\D(x)\|_2 \leq \kappa_\D \|P_\mc{C}(x)\|_2,
\end{equation}
where:
\begin{equation}
    \kappa_\D = \left\{ \begin{array}{ll}
        1 &  \mbox{if $\mathcal{D}$ is convex}\\
       2 &  \mbox{if $\mathcal{D}$ is non-convex}
        \end{array}
        \right.
    \end{equation}
where $\D$ is a potentially non-convex closed set included by cone $\mc{C}$. On the other hand, utilizing a simplified result of \citep{pmlr-v97-qian19b} with partition minibatches, we have:
\begin{equation}
\begin{split}
     &\mathbb{E}_S(\|A^TM^TMA(x - z)\|_2^2) \\&\leq 2 L_s (\frac{q^2}{2n}\|Ax - b\|_2^2 - \frac{q^2}{2n}\|Az- b\|_2^2 - q^2 \langle \triangledown f(z), x-z \rangle).
\end{split}
\end{equation}
where $\triangledown f(z) = \frac{1}{n} A^T(Az - b)$. Then for the case of noisy measurements $b= Ax^\dagger + w$, following similar procedure we can have:
\begin{eqnarray*}\label{esmooth}
&&\mathbb{E}_S(\|A^TM^TMA(x - z)\|_2^2)\\ && \leq 2 L_s (\frac{q^2}{2n}\|Ax - b\|_2^2 - \frac{q^2}{2n}\|Az- b\|_2^2 \\&&- q^2 \langle  \frac{1}{n} A^T(Az - b), x-z \rangle)\\
&& \leq \frac{q^2L_s}{n}(\frac{1}{2}\|A(x - x^\dagger) - w\|_2^2 - \|w\|_2^2 + \langle w, A(x- x^\dagger)\rangle)\\
&&=\frac{q^2L_s}{n}\|A(x-x^\dagger)\|_2^2
\end{eqnarray*}

As shown in the theorem, we have assumed the approximation errors of the forward and adjoint operator are bounded:
\begin{equation}
\begin{split}
        &\|M_iA_{s_k}\|_2\|M_iA_{s_k} \D(x_k) - M_iAx_k\|_2 \leq \ve_1 \\
        &\|\U(M_iA_{s_k})^Ty_k - (M_iA)^Ty_k\|_2 \leq \ve_2, \forall i, k
\end{split}
\end{equation}
Denoting:
\begin{eqnarray*}
    y_k := M_iA_{s_k} \D(x_k) - M_i b,
\end{eqnarray*}
then for $k$-th iteration of PnP-MS2G we have the following:
\begin{eqnarray*}
&&\|x_{k+1}-x^\dagger\|_2\\
&=& \|\mathcal{P}_{\theta_p}(x_k- \tau \U{((M_iA_{s_k})}^Ty_k)-x^\dagger\|_2\\
&\leq& \|P_\M(x_k- \tau \U{((M_iA_{s_k)}}^Ty_k)))-x^\dagger\|_2 \\&&+ \|e(x_k- \tau \U{((M_iA_{s_k})}^Ty_k)))\|_2\\
&=& \|P_{\M-x^\dagger}(x_k-x^\dagger- \tau\U{((M_iA_{s_k})}^Ty_k)))\|_2 + \|e(\Bar{x_k})\|_2,
\end{eqnarray*}
Then  we can continue:
\begin{eqnarray*}
&&\|x_{k+1}-x^\dagger\|_2\\ 
&\leq& \kappa\|P_{\mathcal{C}}(x_k-x^\dagger - \tau \U((M_i A_{s_k})^Ty_{k+1}))\|_2 + \|e(\Bar{x_k})\|_2\\ 
&=& \kappa\sup_{v \in \mathcal{C} \cap \mathcal{B}^{d}} v^T(x_k-x^\dagger - \tau \U((M_i A_{s_k})^Ty_{k+1}) + \|e(\Bar{x_k})\|_2\\
&\leq& \kappa\sup_{v \in \mathcal{C} \cap \mathcal{B}^{d}} v^T(x_k-x^\dagger - \tau A^TM_i^T(M_iAx_k-M_ib)) +\varepsilon\\
&=& \kappa\sup_{v \in \mathcal{C} \cap \mathcal{B}^{d}} v^T[x_k-x^\dagger 
\\&&- \tau A^TM_i^T(M_iAx_k-M_i(Ax^\dagger + w))] + \varepsilon\\
&\leq& \kappa\sup_{v \in \mathcal{C} \cap \mathcal{B}^{d}} v^T[(I - \tau A^T{M_i}^T{M_i}A)(x_k-x^\dagger)] \\&&+ 2\tau\sup_{v \in \mathcal{C} \cap \mathcal{B}^{d}} v^TA^TM_i^TM_iw + \varepsilon\\
&\leq& \kappa\|(I - \tau A^T{M_i}^T{M_i}A)(x_k-x^\dagger)\|_2 \\&& + 2\tau\sup_{v \in \mathcal{C} \cap \mathcal{B}^{d}} v^TA^TM_i^TM_iw + \varepsilon.
\end{eqnarray*}

Denote $\Bar{x_k}:= x_k- \tau A^T{M_i}^Ty_k$, and take expectation, then we have:
\begin{eqnarray*}
&&\E(\|x_{k+1}-x^\dagger\|_2)\\
&\leq& \kappa\E(\|(I - \tau A^T{M_i}^T{M_i}A)(x_k-x^\dagger)\|_2) \\&&+ \varepsilon \\&& + 2\tau\E\sup_{v \in \mathcal{C} \cap \mathcal{B}^{d}} v^TA^TM_i^TM_iw\\
&\leq& \kappa\sqrt{ \E(\|(I - \tau A^T{M_i}^T{M_i}A)(x_k-x^\dagger)\|_2^2)} \\&&+ \varepsilon \\&&+ 2\tau\E\sup_{v \in \mathcal{C} \cap \mathcal{B}^{d}} v^TA^TM_i^TM_iw\\
&=& \kappa\{\E( \|x_k-x^\dagger\|_2^2-2\tau\|{M_i}A(x_k-x^\dagger)\|_2^2\\&&+\tau^2\|A^T{M_i}^T{M_i}A(x_k-x^\dagger)\|_2^2)\}^{\frac{1}{2}}\\&&+ \varepsilon + 2\tau\E\sup_{v \in \mathcal{C} \cap \mathcal{B}^{d}} v^TA^TM_i^TM_iw\\
\end{eqnarray*}
Now denoting:
\begin{equation}
    \delta:= 2\tau\E\sup_{v \in \mathcal{C} \cap \mathcal{B}^{d}, i \in [m]} v^TA^TM_i^TM_iw
\end{equation}
and then:
\begin{eqnarray*}
&&\E(\|x_{k+1}-x^\dagger\|_2)\\
&\leq& \kappa\{  \|x_k-x^\dagger\|_2^2-2\frac{\tau qv_b}{n}\|A(x_k-x^\dagger)\|_2^2\\&&+\frac{\tau^2q^2L_s v_a}{n} \|A(x_k - x^\dagger)\|_2^2 \}^{\frac{1}{2}}+ \varepsilon + \delta\\
&\leq& \kappa\{  \|x_k-x^\dagger\|_2^2-(2\tau q v_b-2L_s\tau^2 q^2v_a)\cdot \frac{1}{n} \|A(x_k - x^\dagger)\|_2^2\}^{\frac{1}{2}}\\&& + \varepsilon+ \delta
\end{eqnarray*}
and then due to Assumption A.3 the Restricted Eigenvalue Condition we have:
\begin{eqnarray*}
&&\E(\|x_{k+1}-x^\dagger\|_2)\\
&\leq& \kappa\{  \|x_k-x^\dagger\|_2^2-(2\tau q\mu_c v_b - 2L_s \mu_c\tau^2 q^2 v_a) \|x_k - x^\dagger\|_2^2\}^{\frac{1}{2}} \\&& + \varepsilon+ \delta\\
&=&\kappa\{1- 2\mu_c\tau q v_b+ 2L_s \mu_c\tau^2q^2 v_a\}^{\frac{1}{2}}\|x_k-x^\dagger\|_2  \\&&+ \varepsilon+ \delta\\
&\leq& \kappa(1-\frac{\mu_cv_b}{L_sv_a})\|x_k-x^\dagger\|_2  + \varepsilon + \delta
\end{eqnarray*}
Then let $\alpha=\kappa(1-\frac{\mu_cv_b}{L_sv_a})$, by the tower rule we get:
 \begin{eqnarray*}
     \E(\|x_k-x^\dagger\|_2) &\leq& \alpha^{{k}}\|x_0-x^\dagger\|_2 + \frac{(1- \alpha^{k})}{1-\alpha} (\ve + \delta).
 \end{eqnarray*}

\subsection{Proof for Theorem 3.2}

For proving the lower bound we will need to assume the constraint set $\M$ to be convex and apply a know result provide in \citep[Lemma F.1]{oymakbabak2016sharp}, that for a closed convex set $\D:= \M - x^\dagger$ containing the origin, given any $a, \gamma \in (0,1]$ there exist a positive constant $C$ such that for any $v$ satisfies $\|\mc{P}_\mc{C}(v)\|_2 \geq a \|v\|_2$ and $\|v\|_2 \leq c$, we can have:
\begin{equation}\label{lemo}
    \frac{\|P_\D(v)\|_2}{\|P_\mc{C}(v)\|_2} \geq 1 - \gamma.
\end{equation}
Since in A.2 we assume the ground truth $x^\dagger \in \M$  we know that $0 \in \M - x^\dagger$ hence the above claim is applicable.
For $k$-th layer of simplified LSPD we have the following:
\begin{eqnarray*}
&&\|x_{k+1}-x^\dagger\|_2\\
&=& \|\mathcal{P}_{\theta_p}(x_k- \tau \U(A_{s_k}^T{M_i}^Ty_k))-x^\dagger\|_2\\
&\geq& \|P_\M(x_k- \tau \U(A_{s_k}^T{M_i}^Ty_k))-x^\dagger\|_2  \\&&- \|e(x_k- \tau \U(A_{s_k}^T{M_i}^Ty_k))\|_2\\
&=& \|P_{\M-x^\dagger}(x_k-x^\dagger- \tau \U(A_{s_k}^T{M_i}^Ty_k))\|_2 \\&&- \|e(x_k- \tau \U(A_{s_k}^T{M_i}^Ty_k))\|_2.
\end{eqnarray*}
Now due to (\ref{lemo}) we can continue:
\begin{eqnarray*}
&&\|x_{k+1}-x^\dagger\|_2 \\&\geq& (1 - \gamma)\|P_{\mc{C}}(x_k-x^\dagger- \tau \U(A_{s_k}^T{M_i}^Ty_k))\|_2 - \varepsilon_0\\
&=& (1 - \gamma) \sup_{v\in \mc{C} \cap \mc{B}^d} v^T(x_k-x^\dagger- \tau \U(A_{s_k}^T{M_i}^Ty_k)) - \varepsilon_0\\
&=& (1 - \gamma) \sup_{v\in \mc{C} \cap \mc{B}^d} v^T(x_k-x^\dagger- \tau A^T{M_i}^Ty_k) \\&&- \varepsilon_0 - \tau \ve_2\\
&=& (1 - \gamma) \sup_{v\in \mc{C} \cap \mc{B}^d} v^T(x_k-x^\dagger\\&&- \tau A^T{M_i}^T(M_iA_{s_k} \D(x_k) - M_ib)) - \varepsilon_0 - \tau \ve_2\\
&=& (1 - \gamma) \sup_{v\in \mc{C} \cap \mc{B}^d} v^T(x_k-x^\dagger\\&&- \tau A^T{M_i}^T(M_iAx_k - M_ib)) - \varepsilon_0 - \tau \ve_2 - \tau \ve_1\\
&=& (1 - \gamma)\|P_{\mc{C}}[(I- \tau A^T{M_i}^TM_iA)(x_k-x^\dagger)]\|_2 \\&&- \varepsilon\\
&=& (1 - \gamma)\sup_{v \in \mc{C} \cap \mb{S}^{d-1}} v^T(I- \tau A^T{M_i}^TM_iA)(x_k-x^\dagger)- \varepsilon\\
&\geq& (1 - \gamma) \frac{x_k - x^\dagger}{\|x_k - x^\dagger\|_2}(I- \tau A^T{M_i}^TM_iA)(x_k-x^\dagger)- \varepsilon\\
&=& (1 - \gamma) \|x_k - x^\dagger\|_2\left(1 - \tau\frac{\|M_iA(x_k - x^\dagger)\|_2}{\|x_k - x^\dagger\|_2}\right) - \varepsilon\\
\end{eqnarray*}
where we denote $\ve = \varepsilon_0 + \tau \ve_2 + \tau \ve_1$. On the other hand since:
\begin{equation}
    \|(I- \tau A^T{M_i}^TM_iA)(x_k-x^\dagger)\|_2  \leq (1+\tau qL_s)\|(x_k-x^\dagger)\|_2,
\end{equation}
and also note the second part of restricted eigenvalue condition we have:
\begin{equation}
    \|M_iA(x_k - x^\dagger)\|_2 \leq qL_c \|x_k - x^\dagger\|_2.
\end{equation}
Hence:
\begin{eqnarray*}
&&\|P_{\mc{C}}[(I- \tau A^T{M_i}^TM_iA)(x_k-x^\dagger)]\|_2\\
    &\geq& \|x_k - x^\dagger\|_2\left(1 - \tau\frac{\|M_iA(x_k - x^\dagger)\|_2}{\|x_k - x^\dagger\|_2}\right)  \\
    &\geq&  \left(\frac{1 - q\tau L_c}{1+q\tau L_s}\right) \|(I- \tau A^T{M_i}^TM_iA)(x_k-x^\dagger)\|_2
\end{eqnarray*}
Combining these three with $\tau = \frac{1}{q L_s}$, we find that (\ref{lemo}) is satisfied for the choice $v = (I- \tau A^T{M_i}^TM_iA)(x_k-x^\dagger)$ and $a = \frac{L_s - L_c}{2L_s}$, we can write:
\begin{equation}
    \|x_{k+1}-x^\dagger\|_2 \geq (1-\gamma)(1- \frac{L_c}{L_s}) \|x_{k}-x^\dagger\|_2 - \varepsilon,
\end{equation}
for all $\|x_k - x^\dagger\|_2 \leq \frac{\delta}{2}$ and by unfolding the iterations to $x_0$ we finish the proof.



\bibliography{main.bib}

\end{document}